\documentclass[12pt]{iopart}
\usepackage{iopams}
\eqnobysec

\begin{document}

%\input{bihamiltonian-macros.sty}

%my macros for LaTeX fixes

% refs: 
%put `( )' around an equation ref in latex
\def\eqref#1{(\ref{#1})}
\def\eqrefs#1#2{(\ref{#1}) and~(\ref{#2})}
\def\eqsref#1#2{(\ref{#1}) to~(\ref{#2})}
\def\sysref#1#2{(\ref{#1})--(\ref{#2})}

%put `Eq.( )' around an equation ref in latex
\def\Eqref#1{Eq.~(\ref{#1})}
\def\Eqrefs#1#2{Eqs.~(\ref{#1}) and~(\ref{#2})}
\def\Eqsref#1#2{Eqs.~(\ref{#1}) to~(\ref{#2})}
\def\Sysref#1#2{Eqs. (\ref{#1}),~(\ref{#2})}

%put `Sec. ' before a section ref in latex
\def\secref#1{Sec.~\ref{#1}}
\def\secrefs#1#2{Sec.~\ref{#1} and~\ref{#2}}

%put `App. ' before an appendix ref in latex
\def\appref#1{Appendix~\ref{#1}}

%put `Ref. ' before a bibitem ref in latex
\def\Ref#1{Ref.~\cite{#1}}

%use footnote style for a bibitem ref in latex
\def\Cite#1{${\mathstrut}^{\cite{#1}}$}

%put `Table ' before a table ref in latex
\def\tableref#1{Table~\ref{#1}}

%put `Fig. ' before a figure ref in latex
\def\figref#1{Fig.~\ref{#1}}

% fix hyphenations to be 
\hyphenation{Eq Eqs Sec App Ref Fig}

% abbrevs for latex commands:
%equations
\def\EQ{\begin{equation}}
\def\EQs{\begin{eqnarray}}
\def\endEQ{\end{equation}}
\def\endEQs{\end{eqnarray}}

%my macros

\def\fewquad{\qquad\qquad}
\def\severalquad{\qquad\fewquad}
\def\manyquad{\qquad\severalquad}
\def\manymanyquad{\manyquad\manyquad}

\def\downupindices#1#2{{\mathstrut}^{}_{#1}{\mathstrut}_{}^{#2}}
\def\updownindices#1#2{{\mathstrut}_{}^{#1}{\mathstrut}^{}_{#2}}
\def\mixedindices#1#2{{\mathstrut}^{#1}_{#2}}
\def\downindex#1{{\mathstrut}_{#1}}
\def\upindex#1{{\mathstrut}^{#1}}

\def\eqtext#1{\hbox{\rm{#1}}}

\def\hp#1{\hphantom{#1}}

% derivatives

\def\der#1{\partial\downindex{#1}}

\def\D#1{D\downindex{#1}}
\def\nD#1#2{D\mixedindices{#2}{#1}}
\def\Dinv#1{D\mixedindices{-1}{#1}}
\def\perpD#1{D\mixedindices{\perp}{#1}}

\def\covder#1{\nabla\downindex{#1}}

\def\covD#1{{\mathcal D}\downindex{#1}}
\def\covperpD#1{{\mathcal D}\mixedindices{\perp}{#1}}
\def\bigD#1{{\bi D}\downindex{#1}}
\def\bigcovder#1{{\bnabla}\downindex{#1}}

% target and spacetime macros

\def\x#1#2{x\mixedindices{#1}{#2}}

\def\id#1#2{\delta\downupindices{#1}{#2}}

\def\g#1#2{g\downupindices{#2}{#1}} 
\def\flat#1#2{\eta\mixedindices{#1}{#2}} 

\def\vol#1#2{\varepsilon\downupindices{#2}{#1}} 
\def\riem#1#2#3{R(#3)\downupindices{#1}{#2}}
\def\conx#1#2{\Gamma\updownindices{#1}{#2}}
\def\scalcurv{\chi}

\def\cross#1#2{\epsilon\downupindices{#2}{#1}}

\def\gtens{g}
\def\gnorm#1{|{#1}|_\gtens}

\def\itens{J}

% frame macros

\def\T{{\vec T}}
\def\N{{\vec N}}
\def\B{{\vec B}}

\def\e#1{e\downindex{#1}}
\def\w#1#2{w\downupindices{#1}{#2}}

\def\vece{{\bi e}_{\perp}}

\def\q#1#2{q\mixedindices{#1}{#2}}
\def\h#1#2{h\mixedindices{#1}{#2}}
\def\v#1#2{v\mixedindices{#1}{#2}}
\def\htang{h_\parallel}
\def\hperp#1#2{h_\perp\mixedindices{#1}{#2}}
\def\wtang#1#2{\Omega\downupindices{#1}{#2}}
\def\wflow#1#2{\Theta\downupindices{#1}{#2}}

% curve macros

\def\map{\gamma}
\def\mapder#1{\gamma\downindex{#1}}

\def\curv{\kappa}
\def\tors{\tau}
\def\curvder#1{\curv_{#1}}
\def\torsder#1{\tors_{#1}}

\def\tangvec{X}
\def\flowvec{Y}

\def\Cmap{\Gamma}

% PDE macros

%\def\uvec{{\bi u}}

\def\u{{\vec u}}
\def\uder#1{\u\downindex{#1}}

\def\nls{\phi}
\def\nlsder#1{\nls\downindex{#1}}
\def\sg{\theta}
\def\sgder#1{\sg\downindex{#1}}
\def\mkdv{\nu}
\def\mkdvder#1{\mkdv\downindex{#1}}

\def\vecmkdv{\bi{v}}
\def\vecmkdvder#1{\vecmkdv\downindex{#1}}
\def\vecsg{\bi{k}}
\def\vecsgder#1{\vecsg_{#1}}
\def\tangsg{k}
\def\tangsgder#1{\tangsg_{#1}}

% Hamiltonian macros

\def\Rop{{\mathcal R}}
\def\Jop{{\mathcal J}}
\def\Hop{{\mathcal H}}
\def\Iop{{\mathcal I}}

\def\Ham#1{H\upindex{#1}}
\def\funct#1{\mathfrak{#1}}

\def\sympvar#1#2{h_\perp\mixedindices{#1}{#2}}
\def\cosympvar#1#2{\varpi\mixedindices{#1}{#2}}
\def\flowvar{v}
\def\auxcosympvar#1#2{\varpi'\mixedindices{#1}{#2}}
\def\auxsympvar#1#2{h_\perp'\mixedindices{#1}{#2}}

\def\Hfield#1{\partial/\partial{#1}}
\def\coHfield#1{d{#1}}

\def\cosympvec{{\vec \cosympvar{}{}}}
\def\sympvec{{\vec \sympvar{}{}}}
\def\qvec{{\vec \q{}{}}}
\def\hvec{{\vec \h{}{}}}
\def\qvecder#1{{\vec \q{}{}}\downindex{#1}}

\def\auxcosympvec{{\vec \auxcosympvar{}{}}}
\def\auxsympvec{{\vec \auxsympvar{}{}}}

\def\vredvec{\bi{ \v{}{} }}
\def\vredvecder#1{\bi{ \v{}{}}\downindex{#1}}
\def\cosympredvec{\boldsymbol{ \cosympvar{}{}}}
\def\sympredvec{\bi{ \sympvar{}{}}}
\def\auxcosympredvec{\boldsymbol{ \cosympvar{}{}}'}
\def\auxsympredvec{\bi{ \auxsympvar{}{}}}

\def\cosymppot{\cosympredvec{}^\perp}
\def\auxcosymppot{\auxcosympredvec{}^\perp}
\def\symppot{\bi{\h{}{}}^\perp}
\def\unitvec{{\hat n}}

\def\Csympvec{{\vec \h{}{}}_C}
\def\perpCsympvec{{\vec \h{}{}}_{C^\perp}}
\def\perpCcosympvec{{\vec \cosympvar{}{}}_{C^\perp}}
\def\perpCauxcosympvec{{\vec \cosympvar{}{}}'_{C^\perp}}
\def\hC#1{\h{}{#1}}
\def\perpqvec{\qvec^\perp}
\def\perpqvecder#1{\qvec\mixedindices{\perp}{#1}}

\def\hook{\lrcorner}
\def\idop{{\rm id}}

% lie algebra macros

\def\c#1#2{c\downupindices{#1}{#2}}
\def\adq{{\rm ad}(\q{}{})}
\def\adqinv{{\rm ad}(\q{}{})_\perp^{-1}}

\def\cvec#1{{\vec c}\,\upindex{#1}}
\def\ccovec#1{{\vec c}\downindex{#1}}
\def\cvecder#1#2{{\vec c}\mixedindices{\,#1}{#2}}

\def\adop#1{{\rm ad}(#1)}
\def\adinvop#1{{\rm ad}(#1)_\perp^{-1}}
\def\Cq{C_q}
\def\perpCq{C_q^\perp}
\def\Cperpq{C_{\perp q}}

\def\Pop{{\mathcal P}_{\Cq}}
\def\perpPop{{\mathcal P}_{\Cq}^\perp}
\def\Pperpop{{\mathcal P}_{\Cperpq}}

% misc

\def\ie/{i.e.}
\def\eg/{e.g.}
\def\etc/{etc.}
\def\const{{\rm const}}

\def\Rnum#1{{\mathbb R}^{#1}}
\def\Cnum{{\mathbb C}}
\def\i{{\rm i}}
\def\sgn{{\rm sgn}}
\def\inv{{}^{-1}}
\def\invsq{{}^{-2}}

\def\su#1{\mathfrak{su}(#1)}
\def\so#1{\mathfrak{so}(#1)}

\def\O#1{O(#1)}
\def\SO#1{SO(#1)}
\def\SU#1{SU(#1)}
\def\Sp#1{S^{#1}}
\def\Hy#1{H^{#1}}

\def\G{\mathcal{G}}
\def\liealg{\mathfrak{g}}

\def\btimes{{\boldsymbol\times}}

% abrevs
\def\on/{orthonormal}

\title[]
{ Bi-Hamiltonian operators, integrable flows of curves using moving frames, 
and geometric map equations }

\author{ Stephen C. Anco }

\address{
Department of Mathematics\\
Brock University, 
St. Catharines, ON Canada }

\ead{sanco@brocku.ca}

\begin{abstract}
Moving frames of various kinds
are used to derive bi-Hamiltonian operators
and associated hierarchies of multi-component soliton equations
from group-invariant flows of non-stretching curves in
constant curvature manifolds and Lie group manifolds. 
The hierarchy in the constant-curvature case consists of 
a vector mKdV equation coming from a parallel frame, 
a vector potential mKdV equation coming from a covariantly-constant frame,
and higher order counterparts generated by 
an underlying vector mKdV recursion operator.
In the Lie-group case the hierarchy comprises 
a group-invariant analog of the vector NLS equation
coming from a left-invariant frame, 
along with higher order counterparts generated by a recursion operator
that is like a square-root of the mKdV one. 
The corresponding respective curve flows are found to be given by
geometric nonlinear PDEs, specifically mKdV and group-invariant analogs of
Schr\"odinger maps. 
In all cases the hierarchies also contain 
variants of vector sine-Gordon equations arising from 
the kernel of the respective recursion operators. 
The geometric PDEs that describe the corresponding curve flows
are shown to be wave maps. 
Full details of these results are presented for two main cases:
$S^2,S^3\simeq SU(2)$.
\end{abstract}

%\ams{}\pacs{}

\maketitle

\section{ Introduction }

Over the years there has been much interest in the close connection 
between classical soliton equations such as 
Korteweg de Vries (KdV), nonlinear Schr\"odinger (NLS), sine-Gordon (SG), 
on one hand,
and nonlinear geometric partial differential equations (PDEs)
on the other hand,
particularly wave maps and Schr\"odinger maps 
which are natural generalizations of 
the ordinary wave equation and Schr\"odinger equation in $1+1$ dimensions
for a function taking values in a Riemannian or K\"ahler target space $N$.

One prominent example is the $\O{3}$ Heisenberg model 
$\uder{t} = \u\times\uder{xx}$, 
$|\u|=1$, 
which is geometrically just a Schr\"odinger map equation 
$\mapder{t} = \itens\covder{x}\mapder{x}$
where $\map(t,x)$ is a map into the target space $\Sp{2}$,
and $\covder{x}$, $\itens$ are the pull back of the covariant derivative
and the complex structure tensor (Hodge-star operator)
determined by the metric on $\Sp{2}$
in terms of the local coordinates represented by $\map$. 
This model is well known to be equivalent to the NLS equation
$0=\i\nlsder{t} + \nlsder{xx} +2|\nls|^2 \nls$
by means of a transformation \cite{Lakshmanan}
found by Hasimoto. 
\footnote{In fact, the Hasimoto transformation was already known
at the start of the last century \cite{daRios}.}
The transformation is $\nls=\curv\exp(\i\int\tors dx)$
where $\curv,\tors$ are the curvature and torsion associated to $\u$
viewed as the tangent vector 
to an arclength-parametrized 
%helical 
space curve in $\Rnum{3}$, 
whose Euclidean Frenet frame is 
$(\T,\N,\B)=(\u,\curv\inv\uder{x},\curv\inv\u\times\uder{x})$, 
evolving under the Heisenberg equation. 
This evolution gives the equations 
$\curvder{t}
= -2\tors\curvder{x} - \curv\torsder{x}$, 
$\torsder{t} 
= \curv\inv\curvder{xxx} - \curv\invsq\curvder{x}\curvder{xx}
+\curv\curvder{x} - 2\tors\torsder{x}$,
coinciding with the motion of a vortex filament
studied by Hasimoto \cite{Hasimoto}. 
A more direct geometrical formulation of the Hasimoto transformation
has been obtained in the context of recent analytical work \cite{Shatah}
on the Cauchy problem for $\Sp{2}$ Schr\"odinger maps. 
The main idea is to introduce a covariantly-constant \on/ frame 
$\e{1},\e{2}=\itens\e{1}$ 
satisfying $\covder{x}\e{1}=\covder{x}\e{2}=0$
along the curve in $\Sp{2}$ defined by $\map(t,x)$ for $t=\const$. 
The evolution of the components of the tangent vector
$\mapder{x}=\q{}{1}\e{1}+\q{}{2}\e{2}$
in this frame 
then reduces precisely to the NLS equation for $\nls=\q{}{1}+\i\q{}{2}$. 

Another example of wide interest is the $\O{3}$ sigma model
$\uder{tx} = -\u \uder{t}\cdot\uder{x}$, 
$|\u|=1$. 
This model has long been known to be equivalent \cite{Polhmeyer}
to the SG equation
$0=\sgder{tx}+\sin\sg$
through the geometrical relation 
$\uder{t}\cdot\uder{x}
= |\uder{t}||\uder{x}| \cos\sg$
combined with a conformal scaling of $t,x$, 
such that $|\uder{t}|=|\uder{x}|=1$
as allowed by the conservation laws $\D{x}|\uder{t}|=\D{t}|\uder{x}|=0$. 
The equivalence can be looked on as a Hasimoto transformation \cite{Lamb}
$\sg = -\int\tors dx$
in terms of the torsion $\tors$ given by $\u$ 
which is again viewed as determining a Euclidean Frenet frame 
$(\T,\N,\B)=(\u,\curv\inv\uder{x},\curv\inv\u\times\uder{x})$
of an arclength-parameterized 
%helical
space curve in $\Rnum{3}$
but now evolving under the sigma model equation 
and subject to constant curvature $\curv=1$. 
The equations for the evolution of the torsion and curvature are simply
$\curvder{t}=0$, 
$\torsder{tx} = -\tors\sqrt{1-\torsder{t}{}^2}$.
A deeper origin for the SG equation 
comes from the geometric formulation of the $\O{3}$ sigma model 
as an $\Sp{2}$ wave map equation $\covder{x}\mapder{t}=0$
where $\map(t,x)$ is again a map into the target space $\Sp{2}$. 
The curve for $t=\const$ defined by this map $\map(t,x)$ 
moves with uniform speed and does not stretch, 
because of the previous conservation laws. 
Consequently, 
a conformal scaling of $t,x$ allows a covariantly-constant \on/ frame 
$\e{1},\e{2}$ to be adapted to the evolution vector 
$\mapder{t}=\e{1}$,
with $\covder{x}\e{1}=\covder{x}\e{2}=0$
holding along the curve. 
The tangent vector then takes the form 
$\mapder{x}=\cos\sg\e{1}+\sin\sg\e{2}$
in terms of a rotation angle $\sg$
whose evolution is just given by the SG equation. 

From a closely related point of view, 
the mKdV equation is known to arise \cite{DoliwaSantini}
as an $\O{3}$-invariant motion of 
a non-stretching curve $\map(t,x)$ in $\Sp{2}$
as formulated using a moving \on/ frame adapted to the curve
so that 
$\mapder{x}= \curv\e{1}$, 
$J\mapder{x}= \curv\e{2}$, 
(\ie/ an intrinsic Frenet frame)
where $\curvder{t}=0$ is the condition of no stretching. 
$\O{3}$ invariance means that the frame components of the motion 
$\mapder{t}=\h{}{1}\e{1}+\h{}{2}\e{2}$
must depend only on differential invariants of the curve
under the group action determined by the isometry group $\O{3}$ of $\Sp{2}$. 
There is a fundamental invariant given by the 
intrinsic curvature of $\map(t,x)$
defined through the Serret-Frenet equations on $\Sp{2}$ for the moving frame 
$\covder{x} \e{1} = \tors\e{2}$, 
$\covder{x} \e{2} = -\tors\e{1}$, 
with $x$ conformally scaled proportional to the arclength, 
so $\curv=\const$.
This invariant $\tors$ also can be viewed as 
the torsion of 
an arclength-parameterized 
%helical 
space curve in $\Rnum{3}$ 
whose Euclidean Frenet frame is associated to $\map(t,x)$
via the identifications
$\map \mapsto \T=\u$,
$\e{1} \mapsto \N=\curv\inv\uder{x}$,
$\e{2} \mapsto \B=\kappa\inv\u\times\uder{x}$
under the embedding of $\Sp{2}$ into $\Rnum{3}$
given by $|\u|=1$. 
Since the curvature of this space curve in $\Rnum{3}$ is $\curv=\const$, 
a complete set of differential invariants is obtained from 
$x$-derivatives of the torsion $\tors$,
and thus 
$\h{}{1}=\h{}{1}(\tors,\torsder{x},\torsder{xx},\ldots)$, 
$\h{}{2}=\h{}{2}(\tors,\torsder{x},\torsder{xx},\ldots)$. 
The non-stretching condition on the motion of the curve
imposes the relation $\D{x}\h{}{1}= \tors\h{}{2}$,
while the equation of motion of the curve yields the flow
$\torsder{t} = \curv\inv\D{x}( \D{x}\h{}{2} + \tors\h{}{1} ) +\curv \h{}{2}$.
For the particular motion 
$\h{}{1}=\frac{1}{2}\tors^2-\curv^2$, $\h{}{2}= \torsder{x}$, 
this flow becomes precisely the mKdV equation 
$\torsder{t} = \curv\inv(\torsder{xxx}+\frac{3}{2}\tors^2\torsder{x})$
on $\tors$. 
More remarkably, 
as shown in \Ref{DoliwaSantini}, 
when the flow of a non-stretching curve 
is expressed in the operator form 
$\torsder{t} 
= \curv\inv\Rop(\h{}{2}) +\curv \h{}{2}$
with $\h{}{1}=\Dinv{x}(\tors\h{}{2})$ 
put in terms of the normal frame component $\h{}{2}$ of the evolution vector,
it contains the recursion operator 
$\Rop = \nD{x}{2} + \tors^2 + \torsder{x}\Dinv{x}\tors$ 
of the hierarchy of mKdV soliton equations. 
For instance, the 5th order mKdV equation on $\tors$ arises from
the curve motion generated by 
$\h{}{2} 
= \Rop(\torsder{x}) -\curv^2 \torsder{x}
= \torsder{xxx} +(\frac{3}{2}\tors^2 -\curv^2) \torsder{x}$, 
$\h{}{1}
= \Dinv{x}(\tors\Rop(\torsder{x}) -\curv^2 \tors\torsder{x})
= \tors\torsder{xx} -\frac{1}{2} \torsder{x}^2 +\frac{3}{8} \tors^4 
-\frac{1}{2}\curv^2 \tors^2$. 

More generally, other recent work has studied 
group-invariant motions of non-stretching curves $\map(t,x)$ 
in higher dimensional constant curvature spaces 
$N\simeq \Sp{n},\Hy{n},\Rnum{n}$,
first in $n=3$ dimensions \cite{SandersWang1}
and subsequently in all dimensions $n\geq 2$ \cite{SandersWang2}.
The main results give a geometric origin of  
the vector mKdV equation
$\vecmkdvder{t} = 
\vecmkdvder{xxx} + \frac{3}{2}{\vecmkdv\cdot\vecmkdv}\, \vecmkdvder{x}$
and its associated hierarchy of integrable soliton equations,
for an $n-1$-component vector variable $\vecmkdv(t,x)$
given by the frame components of the principal normal along the curve,
generalizing the scalar case $n=2$.
This generalization relies on the use of 
a parallel moving (\on/) frame adapted to the curve $\map(t,x)$,
differing from an intrinsic Frenet (\on/) frame. 
In a parallel frame \cite{Bishop}
the normal frame vectors are defined to have 
a purely tangential derivative along the curve,
while the derivative of the tangent frame vector is normal to the curve. 
These properties uniquely determine a Frenet frame in $n=2$ dimensions; 
in $n>2 $ dimensions a parallel frame is related to a Frenet frame 
by a certain local $\SO{n-1}$ rotation acting on the normal vectors,
and in the case $n=3$ \cite{Wang} this rotation angle is given by 
the formula for a Hasimoto transformation in terms of the torsion 
determined by the Serret-Frenet equations of the frame. 
As shown in \Ref{SandersWang2,AncoWang}, 
the Cartan structure equations 
(which generalize the Serret-Frenet equations \cite{Guggenheimer})
for such a frame and its associated connection matrix 
coming from the flow of a non-stretching curve
can be seen to encode 
compatible Hamiltonian symplectic and cosymplectic operators 
with respect to the Hamiltonian flow variable $\vecmkdv$; 
moreover, these operators produce a hereditary recursion operator 
whenever the flow is invariant 
under the $\SO{n-1}$ structure group preserving the frame. 
In the case $n=2$, 
for spaces $N$ with constant Gaussian curvature $\scalcurv$
(\ie/ $\scalcurv =+1$ for the sphere $N=\Sp{2}$, 
$\scalcurv =-1$ for the hyperboloid $N=\Hy{2}$, 
$\scalcurv =0$ for the plane $N=\Rnum{2}$), 
the encoding of the operators is especially simple. 
Here $\vecmkdv$ is just a $1$-component variable identified with 
the invariant $\tors$ of the curve $\map(t,x)$,
while $x$ is the arclength so $\curv=1$. 
The frame equation of motion of $\map(t,x)$ has the Hamiltonian formulation
$\torsder{t} = \Hop(\cosympvar{}{}) +\scalcurv\Iop(\cosympvar{}{})$
where $\Hop,\Iop$ are compatible mKdV Hamiltonian (cosymplectic) operators,
defined by 
$\cosympvar{}{} = \D{x}\sympvar{}{} +\tors\Dinv{x}(\tors\sympvar{}{}) 
= \Iop\inv(\sympvar{}{})$
which is a Hamiltonian symplectic operator
and $\D{x}\cosympvar{}{}=\Hop(\cosympvar{}{})$.  
In geometrical terms, 
$\sympvar{}{}$ is the normal frame component of the curve motion $\mapder{t}$,
and $\cosympvar{}{}$ is the associated connection component 
in the Serret-Frenet equations for this motion,
$\covder{t}\e{1}=\cosympvar{}{}\e{2}$, 
$\covder{t}\e{2}=-\cosympvar{}{}\e{1}$. 
Viewed in a Hamiltonian setting, 
$\sympvar{}{}\Hfield{\tors}$ and $\cosympvar{}{}\coHfield{\tors}$ 
represent a Hamiltonian vector field and covector field 
on the $x$-jet space of the flow variable $\tors(t,x)$. 
Because $\Hop,\Iop$ are a bi-Hamiltonian pair,
$\Rop=\Hop\circ\Iop\inv$ is a recursion operator 
on Hamiltonian vector fields $\sympvar{}{}\Hfield{\tors}$, 
while its adjoint $\Rop^*=\Iop\inv\circ\Hop$ is a recursion operator
on covector fields $\cosympvar{}{}\coHfield{\tors}$
with $\cosympvar{}{}=\delta \Ham{}/\delta\tors$
holding for some Hamiltonian expressions 
$H=H(\tors,\torsder{x},\torsder{xx},\ldots)$.
This structure gives rise to a hierarchy of
commuting Hamiltonian mKdV flows on $\tors$,
corresponding to integrable non-stretching curve motions in $\Sp{2}$. 

Bringing together all these ideas, 
the purpose of this paper is, 
firstly, to identify the geometric ``map equations''
corresponding to the hierarchy of integrable flows of non-stretching curves 
in constant curvature spaces
and, secondly, to extend this result as well as the geometric relation 
known so far between the parallel moving frame formulation of these flows
and bi-Hamiltonian operators 
to other kinds of frames and more general target spaces. 
The main results will be derived in two and three dimensions. 

In particular, it is shown that:
\vskip0pt
$\bullet$ 
the curve flow in $\Sp{2}$ corresponding to the scalar mKdV equation
in a parallel moving frame is described by an mKdV analog of 
the Schr\"odinger map equation;
\vskip0pt
$\bullet$ 
in a covariantly-constant moving frame
the mKdV curve flow in $\Sp{2}$ corresponds to 
the scalar potential mKdV equation
and encodes bi-Hamiltonian operators yielding 
the mKdV recursion operator in potential form;
\vskip0pt
$\bullet$ 
the kernel of the mKdV recursion operator gives rise to 
the SG equation describing a curve flow in $\Sp{2}$ given by 
the wave map equation;
\vskip0pt
$\bullet$ 
analogous curve flows in $\Sp{3}$ formulated in both 
covariantly-constant and parallel moving frames 
yield bi-Hamiltonian operators associated to 
$\O{2}$-invariant vector generalizations of 
the potential mKdV equation and the SG equation;
\vskip0pt
$\bullet$ 
an enlarged hierarchy of curve flows in $\Sp{3}$
arises through a ``square-root'' of the vector mKdV recursion operator
derived from an encoding of bi-Hamiltonian operators
in a left-invariant moving frame which is tied to the Lie group structure
$\SU{2}\simeq \Sp{3}$;
\vskip0pt
$\bullet$ 
the bottom flow in the enlarged hierarchy is 
a group-invariant analog of the Schr\"odinger map equation. 

In addition, 
higher dimensional generalizations of 
these geometric map equations, vector soliton equations, 
and bi-Hamiltonian operators
will be obtained for flows of curves 
in constant curvature spaces via a covariantly-constant moving frame
and in Lie group spaces via a left-invariant moving frame. 
Underlying all these results is a main insight that 
the torsion and curvature parts of the Cartan structure equations
associated to the general flow of a curve
as formulated by a suitable moving frame 
(adapted either to the curve or to the geometry in which the curve moves)
together carry a geometrical encoding of bi-Hamiltonian operators. 

Related work on derivations of soliton equations from flows of curves
in conformal and similarity geometries, affine geometry, 
and Klein geometries appears in 
\Ref{WangSanders,AncoWang,ChouQu1,ChouQu2,ChouQu3,ChouQu4}.

Key definitions pertaining to Hamiltonian structures
are stated in the appendix.
See \Ref{Dorfman,Olver} for a full summary of Hamiltonian theory 
developed for PDE systems.

\section{ Wave maps and mKdV maps 
from integrable flows of curves on the sphere and the hyperboloid }
%$\Sp{2}$ and $\Hy{2}$ 

Consider a flow of a non-stretching curve $\map(t,x)$
in a two-dimensional constant curvature space: 
$N=\Sp{2}$ the sphere, 
or $N=\Hy{2}$ the hyperboloid,
or $N=\Rnum{2}$ the plane. 
Let $\scalcurv$ denote the Gaussian curvature, 
respectively $+1,-1,0$, 
let $\gtens$ denote the metric tensor,
and let $J$ be the Hodge-star operator. 
To begin, 
the flow equation of motion of $\map(t,x)$ will be shown to 
exhibit a natural Hamiltonian structure when formulated
in a covariantly-constant \on/ frame, 
\EQ
\covder{x}\e{1}=\covder{x}\e{2}=0 ,\quad
\e{2}=J\e{1} . 
\endEQ
First, the non-stretching condition $\gnorm{\mapder{x}}=1$
with $x$ scaled to be the arclength 
implies that 
the tangent vector and normal vector to the curve $\map(t,x)$
are given by 
\EQ 
\mapder{x} = \cos\sg \e{1} +\sin\sg \e{2} = \tangvec ,\quad
J\mapder{x} = \sin\sg \e{1} -\cos\sg \e{2} = J\tangvec , 
\endEQ
where the variable $\sg$ is a rotation angle relating
the covariantly-constant frame to an adapted (Frenet) frame
$\tangvec,J\tangvec$.
Note the Serret-Frenet equations become 
\EQ
\covder{x}\tangvec = \mkdv J\tangvec
\endEQ
which yields the expression 
\EQ
\mkdv = -\sgder{x}
\endEQ
for the intrinsic curvature of $\map(t,x)$ in $N$. 
Hence $\sg = -\int\mkdv dx$ is a nonlocal curvature invariant of $\map(t,x)$. 
Due to this relationship, 
the Hamiltonian structure arising from 
the covariantly-constant frame equations
will be analogous to a potential formulation. 
In this frame
a non-stretching curve motion $\mapder{t} = \h{}{1}\e{1} + \h{}{2}\e{2}$
is $\O{3}$-invariant if, and only if, 
$\h{}{1}=\h{}{1}(\sg,\sgder{x},\sgder{xx},\ldots)$
and 
$\h{}{2}=\h{}{2}(\sg,\sgder{x},\sgder{xx},\ldots)$
are functions of the (differential) invariants of the curve. 
An invariant of motion is 
$\cosympvar{}{}$ coming from the geometrical equation
\EQ
\covder{t}\tangvec = \cosympvar{}{} J\tangvec
\endEQ
for the evolution of the frame. 
In terms of $\sg$
the equation of motion for the curve is given by 
$\Dinv{x}$ applied to the flow equation on $\mkdv$ 
(cf. section~1),
\EQ
\mkdvder{t}= \Rop(\sympvar{}{}) + \scalcurv\sympvar{}{}
\endEQ
which yields
\EQ
-\sgder{t} 
= \Rop^*(\Dinv{x}\sympvar{}{}) + \scalcurv \Dinv{x}\sympvar{}{}
\endEQ
with 
\EQ
\sympvar{}{} =\h{}{1} \sin\sg -\h{}{2} \cos\sg
\endEQ
and
\EQ
\htang = \h{}{1} \cos\sg +\h{}{2} \sin\sg 
= -\Dinv{x}(\sgder{x}\sympvar{}{}) . 
\endEQ
Here
\EQ
\Rop^* = \nD{x}{2} + \mkdv^2 - \mkdv\Dinv{x}\mkdvder{x}
\endEQ
is the adjoint of the mKdV recursion operator
\EQ
\Rop = \nD{x}{2} + \mkdv^2 + \mkdvder{x}\Dinv{x}\mkdv . 
\endEQ
This flow equation on $\sg$ possesses a Hamiltonian form 
\EQ
\label{covfloweq}
-\sgder{t} 
= {\tilde\Hop}(\sympvar{}{}) + \scalcurv{\tilde\Iop}(\sympvar{}{})
\endEQ
where 
\EQ
{\tilde\Hop}(\sympvar{}{}) 
= \D{x}\sympvar{}{} + \sgder{x}\Dinv{x}(\sgder{x}\sympvar{}{})
=\cosympvar{}{} ,\quad 
{\tilde\Iop}(\sympvar{}{}) 
= \Dinv{x}\sympvar{}{}
=\cosympvar{}{}{}' 
\endEQ
are compatible Hamiltonian cosymplectic operators. 
Note $\cosympvar{}{}{}'$ is related to $\cosympvar{}{}$ 
by its inverse image under $\Rop^*$. 
Compared to the formulation in an adapted frame, 
here $\sympvar{}{}$ (normal frame component of motion)
and $\cosympvar{}{}$ (connection component for the frame motion)
switch roles,
so on the $x$-jet space of the flow variable $\sg(t,x)$,
$\cosympvar{}{}\Hfield{\sg}$ represents a Hamiltonian vector field
(as likewise does $\cosympvar{}{}{}'\Hfield{\sg}$)
and $\sympvar{}{}\coHfield{\sg}$ represents a related covector field. 
A corresponding feature is that 
the operators ${\tilde\Iop}$ and ${\tilde\Hop}$
are respective inverses of the operators $\Hop$ and $\Iop$
appearing in the Hamiltonian form of the flow equation on $\mkdv$
(cf. section~1)
\EQ
\mkdvder{t}= \Hop(\cosympvar{}{}) + \scalcurv\Iop(\cosympvar{}{}) .
\endEQ
Consequently
the recursion operators 
$\Rop=\Hop\circ\Iop\inv$
and 
${\tilde\Rop}= {\tilde\Hop}\circ{\tilde\Iop}\inv 
= \Iop\inv\circ\Hop=\Rop^*$
arising in the flow equations on $\mkdv$ and $\sg$
are adjoints. 
These recursion operators generate 
a hierarchy of commuting Hamiltonian flows on $\sg$ and $\mkdv$. 

The structure of the hierarchy looks simplest in the planar case, 
$\scalcurv=0$. 
The flow equation \eqref{covfloweq} 
produces a hierarchy of commuting flows
$-\sgder{t} = \cosympvar{(k)}{} = {\tilde\Iop}(\sympvar{(k)}{})$
generated by the operator ${\tilde\Rop}$, 
with involutive Hamiltonians 
$\Ham{}=\Ham{(k)}$ given by 
$\delta \Ham{}/\delta\sg = \sympvar{(k)}{}$
forming a hierarchy generated through the adjoint operator
${\tilde\Rop}^*$, 
$k=0,1,2,\ldots$; 
equivalently, there is a hierarchy of 
commuting flows 
$\mkdvder{t}=\sympvar{(k)}{}$ 
and corresponding involutive Hamiltonians $\Ham{}=\Ham{(k)}$
such that $\delta \Ham{}/\delta\mkdv = \cosympvar{(k)}{}$, 
satisfying $\sympvar{(k)}{}=\Hop(\cosympvar{(k)}{})$,
$k=0,1,2,\ldots$, 
generated by the operators $\Rop={\tilde\Rop}^*$ and $\Rop^*={\tilde\Rop}$. 
The relation $\mkdv =-\sgder{x}$ induces a one-to-one correspondence
between the hierarchies of flows 
on the Hamiltonian variables $\mkdv$ and $\sg$. 

The hierarchy starts at the $k=0$ flow, 
\EQ
\sympvar{(0)}{}=\mkdvder{x} ,\quad
\cosympvar{(0)}{}=\mkdv ,\quad
\Ham{(0)}=\frac{1}{2}\mkdv^2=\frac{1}{2}\sgder{x}^2 , 
\endEQ
which produces a convective (traveling wave) equation 
$\mkdvder{t} = \mkdvder{x}$
and $\sgder{t}= \sgder{x}$. 
Higher flows in the hierarchy are given by 
\EQ\label{mkdvhierarchy}
\sympvar{(k)}{}=\Rop^k(\mkdvder{x}) ,\quad
\cosympvar{(k)}{}=\Rop^*{}^k(\mkdv) ,\quad
k=1,2,\ldots . 
\endEQ
In particular the $k=+1$ flow is the mKdV equation 
$\mkdvder{t} = \mkdvder{xxx} + \frac{3}{2} \mkdv^2\mkdvder{x}$
or in potential form 
$\sgder{t} = \sgder{xxx} + \frac{1}{2} \sgder{x}^3$. 
Each flow \eqref{mkdvhierarchy} is bi-Hamiltonian, 
since 
$\cosympvar{(k)}{} 
={\tilde\Hop}(\sympvar{(k-1)}{})
={\tilde\Iop}(\sympvar{(k)}{})$
and hence
\EQ
-\sgder{t} 
= {\tilde\Iop}(\delta \Ham{(k)}/\delta\sg) 
= {\tilde\Hop}(\delta \Ham{(k-1)}/\delta\sg) , \quad 
k=1,2,\ldots,
\endEQ
or equivalently
\EQ
\mkdvder{t} 
= \Hop(\delta \Ham{(k)}/\delta\mkdv) 
= \Iop(\delta \Ham{(k+1)}/\delta\mkdv) ,\quad
k=0,1,2,\ldots .
\endEQ

The hierarchy also contains a $k=-1$ flow 
characterized by the property that 
it gets mapped into the stationary flow 
$\sgder{t}=0=\mkdvder{t}$ under the recursion operators 
$\Rop$ and ${\tilde\Rop}$. 
Hence in this flow 
$\sympvar{(-1)}{}$ and $\cosympvar{(-1)}{}$
satisfy the equation
$0={\tilde\Hop}(\sympvar{(-1)}{})$
with 
$\cosympvar{(-1)}{} = {\tilde\Iop}(\sympvar{(-1)}{})$,
producing a nonlocal evolution equation on $\mkdv$
related in potential form on $\sg$ to the SG equation
as shown in more detail later. 

Linear combinations of the flows in this hierarchy
produce commuting bi-Hamiltonian flows 
in the non-planar case, $\scalcurv\neq 0$:
\EQ
\mkdvder{t}
= \sympvar{(k+1)}{} +\scalcurv\sympvar{(k)}{} ,\quad
\delta \Ham{(k,\scalcurv)}/\delta\mkdv 
= \cosympvar{(k+1)}{} +\scalcurv\cosympvar{(k)}{} , 
\endEQ
which will be called the $+k$ flow, 
satisfying the property
\EQ
\mkdvder{t} 
%= \Hop(\delta \Ham{}/\delta\mkdv) 
%= \Iop(\Rop^*(\delta \Ham{}/\delta\mkdv))
%where $k$ is shifted by $+1$ in the latter Hamiltonian.
= \Hop(\delta \Ham{(k,\scalcurv)}/\delta\mkdv) 
= \Iop(\delta \Ham{(k+1,\scalcurv)}/\delta\mkdv)
\endEQ
where the Hamiltonians are given by 
\EQ
\Ham{}=\Ham{(k,\scalcurv)} := \scalcurv\Ham{(k)} +\Ham{(k+1)}  . 
\endEQ
In an equivalent potential form, 
the $+k$ flow is 
\EQ
-\sgder{t}
= \cosympvar{(k+1)}{} +\scalcurv\cosympvar{(k)}{} ,\quad
\delta \Ham{(k,\scalcurv)}/\delta\sg 
= \sympvar{(k+1)}{} +\scalcurv\sympvar{(k)}{} ,
\endEQ
which obeys
\EQ
-\sgder{t} 
%= {\tilde\Iop}(\delta \Ham{}/\delta\sg) 
%= {\tilde\Hop}(\Rop(\delta \Ham{}/\delta\sg))
= {\tilde\Iop}(\delta \Ham{(k,\scalcurv)}/\delta\sg) 
= {\tilde\Hop}(\delta \Ham{(k-1,\scalcurv)}/\delta\sg) .
\endEQ
Independently of $\scalcurv$,
associated to the flows on $\mkdv$ and $\sg$
are a hierarchy of commuting Hamiltonian vector fields
$\sympvar{(k)}{}\Hfield{\mkdv}$
and a hierarchy of involutive variational covector fields
$\cosympvar{(k)}{}\coHfield{\mkdv}$, 
$k=0,1,2,\ldots$, 
with a corresponding potential form given by 
switching $\mkdv$ and $\sg$, 
as well as $\sympvar{(k)}{}$ and $\cosympvar{(k)}{}$. 

Finally, the motion of $\map(t,x)$
determined by these flows 
will now be derived
through the relation $\sympvar{}{}=\sympvar{(k)}{}$ 
producing 
a hierarchy of non-stretching curve motions from the hierarchy of 
commuting Hamiltonian vector fields $\sympvar{(k)}{}\Hfield{\mkdv}$. 
The equation of motion for the flow of the curve $\map(t,x)$
is given by 
\EQ
\mapder{t} = \htang \tangvec + \sympvar{}{} J\tangvec
\endEQ
with 
\EQ
\htang= \Dinv{x}(\mkdv\sympvar{}{}) .
\endEQ
The $0$ flow is produced by 
\EQ
\sympvar{}{}=\mkdvder{x} ,\quad
\htang{}{}=\frac{1}{2}\mkdv^2 ,
\endEQ
which gives the equation 
\EQ
\mapder{t} 
= \frac{1}{2} \mkdv^2 \tangvec + \mkdvder{x} J\tangvec .
\endEQ
Integration by parts and substitution of the Serret-Frenet relations 
\EQ
\tangvec=\mapder{x} ,\quad
\mkdv J\tangvec = \covder{x}\mapder{x} 
\endEQ
then yields
\EQ
\mapder{t} 
= \covder{x}^2\mapder{x} 
+\frac{3}{2}\gnorm{\covder{x}\mapder{x}}^2 \mapder{x} ,\quad
\eqtext{ with }\quad 
\gnorm{\mapder{x}}=1 .
\endEQ
This motion will be called a non-stretching {\it mKdV map} equation. 
The higher flows in the hierarchy give 
analogous higher-order mKdV map equations, 
derived recursively via the Serret-Frenet relations 
and their $x$-derivatives:
$\mkdvder{x} J\tangvec 
= \covder{x}(\mkdv J\tangvec) +\mkdv^2\tangvec
= \covder{x}^2\mapder{x}+ \mkdv^2\mapder{x}$
and so on. 
For instance
the $+1$ flow has
\EQ
\sympvar{}{}=\mkdvder{xxx}+\frac{3}{2}\mkdv^2\mkdvder{x} ,\quad
\htang{}{}=\mkdv\mkdvder{xx} -\frac{1}{2} \mkdvder{x}^2 +\frac{3}{8}\mkdv^4 ,
\endEQ
and so
\EQs
\mapder{t} &&
= ( \mkdv\mkdvder{xx} -\frac{1}{2} \mkdvder{x}^2 +\frac{3}{8}\mkdv^4 )\tangvec 
+ ( \mkdvder{xxx}+\frac{3}{2}\mkdv^2\mkdvder{x} )J\tangvec
\\&&
= \covder{x}( \frac{3}{2}\mkdv^3 J\tangvec + \frac{3}{2}(\mkdv^2)_x \tangvec 
+\covder{x}^2(\mkdv J\tangvec) )
+ ( (\mkdv^2)_{xx} -\frac{5}{2} \mkdvder{x}^2 +\frac{7}{8}\mkdv^4 )\tangvec 
\nonumber
\endEQs
which yields
\EQs
\mapder{t} &&
= \covder{x}^4\mapder{x} 
+\frac{3}{2}\gnorm{\covder{x}\mapder{x}}^2 \covder{x}^2\mapder{x}
+ 3(\gnorm{\covder{x}\mapder{x}}^2)\downindex{x} \mapder{x}
\nonumber\\&&\fewquad
+( \frac{5}{2} (\gnorm{\covder{x}\mapder{x}}^2)\downindex{xx}
- \frac{5}{2} \gnorm{\covder{x}^2\mapder{x}}^2 
+\frac{27}{8} \gnorm{\covder{x}\mapder{x}}^4 )\mapder{x} ,
\endEQs
again with $\gnorm{\mapder{x}}=1$. 
In general the $+k$ flow for $k=0,1,\ldots$ 
corresponds to an mKdV map equation of order $3+2k$. 

The $-1$ flow in contrast comes from the equation 
\EQ
0=\cosympvar{}{}=\D{x}\sympvar{}{}+\mkdv\htang ,
\endEQ
which is directly equivalent to 
$\covder{t}\tangvec = \cosympvar{}{} J\tangvec=0$. 
This motion is just a wave map equation, 
\EQ
\covder{t}\mapder{x}=0 ,
\endEQ
subject to the non-stretching condition 
\EQ
\gnorm{\mapder{x}}=1 . 
\endEQ
Furthermore, $t$ can be conformally scaled so that 
the curve motion has unit speed 
\EQ
\gnorm{\mapder{t}}=1 . 
\endEQ
In the planar case, \ie/ $N=\Rnum{2}$, 
the wave map equation reduces to an ordinary wave equation,
as the flow on $\mkdv=-\sgder{x}$ is stationary 
$\mkdvder{t}= -\sgder{tx} =0$ 
due to $\scalcurv=0$. 
The $-1$ flow equation for $\scalcurv\neq 0$
is $\mkdvder{t} = \scalcurv\sympvar{}{}$ 
where $\mkdv = -\htang\inv \D{x}\sympvar{}{}$
and $\htang= \sqrt{1-\sympvar{2}{}}$,
which gives a hyperbolic scalar PDE 
\EQ
\mkdvder{tx} = -\sgn(\scalcurv) \sqrt{\scalcurv^2 -\mkdvder{t}^2} \mkdv
\endEQ
on $\mkdv$. 
In potential form, these relations imply
$\sympvar{}{}=\sin\sg$, $\htang=\cos\sg$, 
up to a shift in $\sg$
(equivalently, $\h{}{1}=1,\h{}{2}=0$ are constants),
while $\cosympvar{}{}{}'=\Dinv{x}\sin\sg$. 
Hence the $-1$ flow in the non-planar case, 
namely a non-stretching wave map on $N=\Sp{2}$ or $\Hy{2}$, 
is given by 
\EQ
\sgder{tx} = -\scalcurv\sin\sg
\endEQ
which is equivalent to the SG equation. 

In addition to all these Hamiltonian flows of curves, 
there is a trivial flow produced by $\sympvar{}{}=0$, $\htang=1$, 
falling outside the hierarchy. 
This flow yields the curve motion 
$\mapder{t}= \tangvec=\mapder{x}$
which is just a convective (traveling wave) map equation. 

{\bf In summary}:
In two-dimensional constant-curvature spaces
there is a hierarchy of bi-Hamiltonian commuting flows of
non-stretching curves $\map(t,x)$; 
the $0$ flow is described by an mKdV map equation, 
such that the curvature invariant of $\map$ in a Frenet frame
satisfies the mKdV equation to within a convective term,
while $+k$ flow is a higher-order analog. 
The wave map equation describes a $-1$ flow such that 
it is mapped into the stationary flow 
under the recursion operator of the hierarchy,
with the curvature invariant of $\map$ in a covariantly-constant frame
satisfying the SG equation.

\section{ Frame formulations of integrable flows of curves 
in three dimensions }

The results just summarized in two dimensions 
have a natural generalization to three (and higher) dimensions 
formulated in a more general geometric setting. 
Let $(N,\gtens)$ be a three-dimensional Riemannian manifold
and consider a flow of a curve $\map(t,x)$ in $N$. 
Write 
$\tangvec=\mapder{x} = \q{a}{}\e{a}$
for the tangent vector along the curve
and 
$\flowvec=\mapder{t} = \h{a}{}\e{a}$
for the evolution vector of the flow of the curve,
where $\e{a}$ is any \on/ frame defined in the tangent space $T_{\map}N$
on the two-dimensional surface swept out by $\map(t,x)$ in $N$. 
Note the orthonormality of the frame is expressed by 
$\gtens(\e{a},\e{b}) = \id{ab}{}$ (Kronecker symbol); 
hereafter this frame metric will be used to freely raise and lower 
frame indices,
and the summation convention is assumed on all repeated frame indices. 
Write $\covder{}$ for the covariant derivative 
determined by the metric $\gtens$ on $N$, 
and $\gnorm{\tangvec}^2= \gtens(\tangvec,\tangvec)$ 
for the metric norm-squared of $\tangvec$ 
on the tangent space $T_{x}N$. 
Suppose $\map(t,x)$ is non-stretching under the flow, 
so $\gnorm{\tangvec}^2= \q{a}{}\q{}{a} =1$
is unit normalized without loss of generality. 
The frame formulation of covariant derivatives is provided by 
the introduction of connection $1$-forms $\w{ab}{}$ (skew in $ab$)
related to the frame vectors $\e{a}$ 
through the Cartan structure equations as follows
(see also \Ref{Guggenheimer}).
On the surface $T_{\map}N$ the content of the Cartan equations
is that the covariant derivatives 
$\covder{x}$ along the curve and $\covder{t}$ along the flow 
have vanishing torsion 
\EQ
\label{torseq}
\covder{x}\mapder{t} - \covder{t}\mapder{x} 
=[\mapder{x},\mapder{t}] =0
\endEQ
and carry curvature 
\EQ
\label{curveq}
[\covder{x},\covder{t}]
= \riem{}{}{\mapder{x},\mapder{t}}
\endEQ
given by the Riemann tensor $\riem{}{}{\cdot,\cdot}$
\cite{KobayashiNomizu} determined from $\gtens$. 
When expressed in frame components, 
the torsion and curvature equations look like 
\EQ\label{frametorseq}
0=\D{x}\h{a}{} - \D{t}\q{a}{} + \h{b}{}\wtang{b}{a} - \q{b}{}\wflow{b}{a}
\endEQ
and 
\EQ\label{framecurveq}
\riem{a}{b}{\mapder{x},\mapder{t}} = 
\D{t}\wtang{a}{b} - \D{x}\wflow{a}{b} 
+\wtang{a}{c}\wflow{c}{b} - \wflow{a}{c}\wtang{c}{b}
\endEQ
where 
\EQ
\wtang{ab}{}
= \mapder{x} \hook \w{ab}{}
= \gtens(\e{b},\covder{x}\e{a})
\endEQ
and 
\EQ
\wflow{ab}{}
= \mapder{t} \hook \w{ab}{}
= \gtens(\e{b},\covder{t}\e{a})
\endEQ
are the Cartan connection matrices, 
and where
\EQ\label{frameriem}
\riem{a}{b}{\mapder{x},\mapder{t}}
= \q{c}{}\h{d}{}\riem{a}{b}{\e{c},\e{d}}
\endEQ
is the Cartan curvature matrix. 
(Alternatively, the connection matrices are seen to arise
from the pullback to $T_{\map}N$ of the equation \cite{KobayashiNomizu}
$\covder{}\e{a} = -\w{a}{b} \otimes \e{b}$
defining a Riemannian connection $\w{ab}{}$ in terms of 
an \on/ frame $\e{a}$ associated with the Riemannian structure
$g$ and $\covder{}$ on $N$.)

These frame equations \eqsref{frametorseq}{frameriem}
will be demonstrated to directly encode
pairs of Hamiltonian operators with respect to some Hamiltonian flow variable
producing a hierarchy of integrable curve flows
if $\e{a}$ and $\w{ab}{}$ are chosen in a geometrical fashion 
determined entirely by the curve $\map$ and the geometry $(N,\gtens)$,
in particular so the frame curvature matrix $\riem{a}{b}{\e{c},\e{d}}$
is everywhere constant on $N$. 

Constant curvature spaces $(N,\gtens)$ have the distinguishing property 
that, for all choices of \on/ frame $\e{a}$,
\EQ
\riem{a}{b}{\e{c},\e{d}} = 2\scalcurv \id{a[c}{}\id{d]}{b}
\quad\eqtext{ with }\quad
\chi=\const ,
\endEQ
namely such spaces are homogeneous and isotropic, 
\ie/ $3$-sphere $N=\Sp{3}$, $\scalcurv=1$;
$3$-hyperboloid $N=\Hy{3}$, $\scalcurv=-1$;
and Euclidean space $N=\Rnum{3}$, $\scalcurv=0$. 
Correspondingly, the local frame structure group is $\SO{3}$. 
Thus, as the frame curvature matrix is constant on $N$,
there is a wide freedom available in determining both
$\e{a}$ and $\w{ab}{}$ geometrically in terms of $\map$ and $\gtens$
on these spaces. 
The precise nature of the encoding of Hamiltonian operators
in the frame structure equations depends essentially on whether
the frame $\e{a}$ is adapted to the tangent vector $\tangvec$ of the curve. 

\subsection{ Parallel frames 
and vector mKdV Hamiltonian operators }

To begin, for the situation of an adapted moving frame $\e{a}$, 
the components of $\tangvec$ are given by constants $\q{a}{}$. 
The Hamiltonian flow variable in this case will be 
the frame components $\v{a}{}$ of $\flowvar= \covder{x}\tangvec$,
with the associated Hamiltonian vector and covector field variables 
$\sympvar{a}{}$, $\cosympvar{a}{}$
given by the frame components of 
$\flowvec_\perp = \flowvec - \gtens(\flowvec,\tangvec)\tangvec$,
$\cosympvar{}{} = \covder{t}\tangvec$. 
The flow equation of motion on $\flowvar$ comes from 
the part of the curvature structure equation \eqref{curveq} projected 
along the tangent direction of $\map$,
$\covder{t}\flowvar 
= \covder{x}\cosympvar{}{} + \scalcurv \flowvec_\perp$
which makes use of the constant curvature property of $(N,\gtens)$. 
The remaining part of the curvature structure equation 
(projected into the normal space of $\map$)
combined with the torsion structure equation \eqref{torseq}
allow $\covder{t},\covder{x},\flowvec_\parallel=\gtens(\flowvec,\tangvec)$
to be expressed in terms of $\cosympvar{}{},\flowvec_\perp$. 
These expressions look simplest if the connection $\w{ab}{}$ is also
adapted to the curve by putting $(\covder{x}(\e{a})_\perp)_\perp=0$, 
in which case $\e{a}$ becomes precisely a parallel frame \cite{Bishop}
such that $\covder{x}(\e{a})_\perp = -\v{}{a}\tangvec$. 
In detail:
$\wtang{ab}{} = 2\q{}{[a}\v{}{b]}$
is the connection equation of the parallel frame;
the torsion equation yields
\EQ
\cosympvar{a}{} = \htang \v{a}{} + \D{x}\hperp{a}{} ,\quad
\D{x}\htang = \hperp{a}{}\v{}{a}
\endEQ
where $\htang=\h{a}{}\q{}{a}$, 
and the normal part of the curvature equation yields
\EQ
\D{x}\wflow{ab}{} 
= 2\cosympvar{}{[a}\v{}{b]} + 2\q{}{[a}\D{x}\cosympvar{}{b]} .
\endEQ
The tangential part of the curvature equation then gives
\EQ
\v{a}{t} = 
\D{x}\cosympvar{a}{} + 2\v{}{b}\Dinv{x}(\cosympvar{[a}{}\v{b]}{})
+ \scalcurv \hperp{a}{} .
\endEQ
Because of the non-stretching condition on $\tangvec$, 
observe 
$\v{a}{}\q{}{a} = \cosympvar{a}{}\q{}{a} =0$
and likewise $\hperp{a}{}\q{}{a}=0$. 
It is now convenient to regard $\v{a}{},\cosympvar{a}{},\hperp{a}{}$
as $2$-component vectors $\vredvec,\cosympredvec,\sympredvec$
and employ an index-free vector notation 
for writing out the Hamiltonian operators. 
Hence, the flow equation is given by 
\EQ
\vredvecder{t} = \Hop(\cosympredvec) + \scalcurv \Jop\inv(\cosympredvec)
= \Rop(\sympredvec) +\scalcurv\sympredvec
\endEQ
in terms of the operators 
\EQ
\label{parframe-HJops}
\Hop = \D{x} + {*\vredvec} \Dinv{x}({*\vredvec}\cdot\ ) ,\quad
\Jop = \D{x} + \vredvec \Dinv{x}(\vredvec\cdot\ ) ,\quad
\Rop = \Hop\circ\Jop ,
\endEQ
where $*$ is the Hodge-star operator on vectors in two dimensions
obeying the product identity
\footnote[2]{
Throughout, in any dimension, 
$\hook$ denotes the interior product between 
a vector and a $1$-form or $2$-form, 
as well as the contraction of a vector against an antisymmetric tensor;
$\wedge$ denotes the exterior product taking a pair of vectors into 
an antisymmetric tensor, 
as well as the ordinary wedge product between forms;
the Hodge-star operator $*$ is mapping between
a vector and a rank $n-1$ skew tensor in $n>2$ dimensions
or between vectors in the $n=2$ dimensional case, with $*^2=-1$.
In addition, vectors and $1$-forms will be identified using the frame metric.
}
$\bi{C}\hook(\bi{B}\wedge\bi{A}) = (*\bi{A}\cdot\bi{B})\, {*\bi{C}}$. 
As proved in \Ref{SandersWang2}
using the framework developed in \Ref{Olver,Dorfman}, 
$\Hop$ and $\Jop$ are a pair of cosymplectic and symplectic 
Hamiltonian operators,
and the formal inverse $\Jop\inv=\Iop$ 
as defined on the $x$-jet space of $\vredvec(t,x)$
is a cosymplectic operator compatible with $\Hop$. 
Moreover, $\Rop$ is a hereditary recursion operator
for a hierarchy of Hamiltonians flows 
that are each invariant under the frame structure group. 
Group-invariance requires the flow components to be 
a covariant vector function
$\sympredvec=\sympredvec(\vredvec,\vredvecder{x},\vredvecder{xx},\ldots)$
on the jet space,
so accordingly the flow equation has an $\O{2}$-invariant vector form,
where this group $\O{2}$ is the isotropy subgroup of 
the frame structure group leaving invariant $\tangvec$
for an adapted (parallel) frame. 
Because a parallel frame is determined geometrically (by the curve $\map$)
up to a rigid $\SO{2}$ rotation,
the Hamiltonian variable $\vredvec$ 
represents differential covariants of $\map$
\cite{MariBeffa}. 

The $0$ flow in the hierarchy is produced by 
$\sympredvec = \vredvecder{x}$, 
giving the vector PDE 
\EQ
\vredvecder{t}= 
\vredvecder{xxx} +\frac{3}{2}|\vredvec|^2\vredvecder{x} 
+\scalcurv \vredvecder{x} .
\endEQ
This flow is a well-known vector generalization \cite{SokolovWolf}
of the mKdV equation,
to within a convective term that can be absorbed by a Galilean transformation
$x\rightarrow x-\scalcurv t$, $t\rightarrow t$. 
The $+k$ flow as obtained from 
$\sympredvec = \Rop^k(\vredvecder{x})$
gives a vector mKdV equation of higher-order $3+2k$ on $\vredvec$. 
These flows correspond to geometric motions of the curve $\map$, 
\EQ\label{curvemotion}
\mapder{t}
= f(\mapder{x},\covder{x}\mapder{x},\covder{x}^2\mapder{x},\ldots)
=\h{a}{}\e{a} 
\endEQ
subject to the non-stretching condition $\gnorm{\mapder{x}}=1$. 
The equation of motion is obtained from the decompositions
$\e{a} =\q{}{a}\tangvec +(\e{a})_\perp$
and $\h{a}{} = \htang\q{a}{} + \hperp{a}{}$
with $\htang=\Dinv{x}(\hperp{a}{}\v{}{a})$,
after substitution of the expressions
$\hperp{a}{} = \hperp{a}{}(\v{b}{},\v{b}{x},\v{b}{xx},\ldots)$
followed by use of the relations 
\EQs
\v{a}{}(\e{a})_\perp = \covder{x}\mapder{x} ,\quad
\v{a}{x}(\e{a})_\perp &&
= \covder{x}(\v{a}{}\e{a}) - \v{a}{}\covder{x}\e{a}
\\&&
= \covder{x}^2\mapder{x} +\gnorm{\covder{x}\mapder{x}}^2 \mapder{x} ,
\nonumber
\endEQs
and so on, 
where 
\EQ
\covder{x}(\e{a})_\perp = -\v{}{a}\mapder{x} . 
\endEQ
In particular, 
for the $0$ flow, 
$\hperp{a}{} =\v{a}{x}$, 
$\htang=\frac{1}{2} \v{a}{}\v{}{a}$, 
thus
\EQ
\mapder{t} = \h{a}{}\e{a}
= \frac{1}{2} \v{a}{}\v{}{a} \tangvec +\v{a}{x} (\e{a})_\perp
= \covder{x}^2\mapder{x} 
+\frac{3}{2}\gnorm{\covder{x}\mapder{x}}^2 \mapder{x} . 
\endEQ
Note the trivial flow given by $\sympredvec=0,\htang=1$ 
corresponds to the motion $\mapder{t}=\mapder{x}$
which is a convective (traveling wave) map equation. 

There is also a $-1$ flow contained in the hierarchy,
with the property that $\sympredvec$ is annihilated by 
the symplectic operator $\Jop$
and hence gets mapped into $\sympredvec=0$ 
under the recursion operator $\Rop$. 
Geometrically this flow means simply $\Jop(\sympredvec)=\cosympredvec=0$
so $\map$ satisfies the equation of motion 
\EQ
\cosympvar{}{} = \covder{t}\mapder{x}=0 
\endEQ
which is the wave map equation on the target space $N=\Sp{3},\Hy{3},\Rnum{3}$. 
The resulting vector PDE is a nonlocal evolution equation 
\EQ
\vredvecder{t} = \scalcurv \sympredvec
\endEQ
where $\sympredvec$ satisfies
\EQ
0=\cosympredvec = \D{x}\sympredvec + \htang \vredvec
\quad\eqtext{ and }\quad
\D{x}\htang = \sympredvec\cdot\vredvec . 
\endEQ
Note this PDE possesses the conservation law 
$0=\D{x}( \htang{}^2+|\sympredvec|^2) = \covder{x}\gnorm{\mapder{t}}^2$
and so a conformal scaling of $t$ can be used to put 
\EQ
1=\gnorm{\mapder{t}}^2= \htang{}^2+|\sympredvec|^2 .
\endEQ
This relation enables $\htang,\sympredvec$ to be expressed 
entirely in terms of $\vredvec$. 
Its geometrical meaning is that the flow is conformally equivalent to
one with uniform speed. 
Consequently, the $-1$ flow vector PDE reduces to 
\EQ
\vredvecder{t} = 
-\sgn(\scalcurv) \Dinv{x}(\sqrt{\scalcurv^2-|\vredvecder{t}|^2} \vredvec)
\endEQ
which is equivalent to a hyperbolic vector equation
\EQ
\vredvecder{tx} = -\sqrt{1-|\vredvecder{t}|^2} \vredvec
\endEQ
after a further scaling $t\rightarrow \scalcurv\inv t$.
This PDE is related to \cite{AncoWolf}
a vector SG equation known from $\O{4}$ sigma models, 
which will be derived later in potential form. 

{\bf Theorem~1}:\ {\it
In three dimensions there is a hierarchy of 
group-invariant integrable flows of curves $\map(t,x)$
in any constant-curvature space 
$(N,\gtens)\simeq \Sp{3},\Hy{3},\Rnum{3}$. 
The $0$ flow is a non-stretching mKdV map equation 
\EQ
\label{mkdvmapeq}
\mapder{t} 
= \covder{x}^2\mapder{x} 
+\frac{3}{2}\gnorm{\covder{x}\mapder{x}}^2 \mapder{x} ,\quad 
\gnorm{\mapder{x}}=1 , 
\endEQ
while the $+k$ flow is a higher-order analog
$\mapder{t} 
=f(\mapder{x},\covder{x}\mapder{x},\ldots,\covder{x}^{2k+2}\mapder{x})$. 
The $-1$ flow is a non-stretching wave map equation
\EQ
\label{wavemapeq}
\covder{t}\mapder{x}=0 ,\quad
\gnorm{\mapder{x}}= \gnorm{\mapder{t}}=1,
\endEQ
which is equivalent to the $\O{4}$ sigma model. 
}

Group-invariance has the geometric meaning here of 
covariance of the equation of motion \eqref{curvemotion} of $\map(t,x)$
under the group of isometries of $(N,\gtens)$,
as implied by the $\O{2}$-invariance property of the vector expressions 
$\sympredvec = \sympredvec(\vredvec,\vredvecder{x},\vredvecder{xx},\ldots)$
on the $x$-jet space of $\vredvec(t,x)$. 
Note these expressions for each $+k$ flow, $k=0,1,\ldots$, 
are explicit polynomials, 
while the $-1$ flow is only given by an implicit nonlocal expression 
in this setting. 

Similarly to the derivation in two dimensions, 
associated to the curve flows in three dimensions is a hierarchy of
commuting Hamiltonian vector fields 
$\sympredvec^{(k)}\cdot\Hfield{\vredvec}$
and involutive variational covector fields 
$\cosympredvec^{(k)}\cdot\coHfield{\vredvec}$,
$k=0,1,2,\ldots$,
generated by the recursion operator $\Rop$ and its adjoint $\Rop^*$.
The hierarchy starts from 
$\sympredvec^{(0)} = \vredvecder{x}$, 
$\cosympredvec^{(0)} = \vredvec$, 
which has the geometrical meaning that the Hamiltonian vector field
$\sympredvec^{(0)}\cdot\Hfield{\vredvec}$ 
is the infinitesimal generator of $x$-translations,
where $x$ is the arclength along the curve $\map$. 
This hierarchy has a natural bi-Hamiltonian structure
$\sympredvec^{(k)} 
= \Hop(\delta \Ham{(k)}/\delta\vredvec)
= \Iop(\delta \Ham{(k+1)}/\delta\vredvec)$
with 
$\cosympredvec^{(k)} = \delta \Ham{(k)}/\delta\vredvec$
determining the Hamiltonians
$\Ham{}=\Ham{}(\vredvec,\vredvecder{x},\vredvecder{xx},\ldots)$. 
Consequently, the curve flows themselves are bi-Hamiltonian:
\EQ
\vredvecder{t} 
= \Hop(\delta \Ham{(k,\scalcurv)}/\delta\vredvec)
= \Iop(\delta \Ham{(k+1,\scalcurv)}/\delta\vredvec)
\endEQ
where 
\EQ
\Ham{(k,\scalcurv)}= \scalcurv\Ham{(k)} +\Ham{(k+1)} .
\endEQ

\subsection{ Covariantly-constant frames 
and vector potential mKdV Hamiltonian operators }

The previous results will now be extended to 
the situation of a non-adapted moving frame $\e{a}$
for curve flows in the same constant-curvature spaces. 
In this case 
the frame components $\q{a}{}$ of the tangent vector $\tangvec$ 
are (non-constant) functions of $t,x$,
and as a consequence the most natural Hamiltonian variable will be given by 
$\flowvar=\tangvec$ itself. 
Compared to the previous case when $\flowvar=\covder{x}\tangvec$,
the roles of the associated variables 
$\flowvec_\perp = \flowvec - \flowvec_\parallel\tangvec$,
$\cosympvar{}{} = \covder{t}\tangvec$
are switched here to be 
Hamiltonian covector and vector field variables, respectively. 
This relationship is analogous to going to a potential formulation 
in the Hamiltonian setting, similarly to the situation in two dimensions. 
The flow equation of motion on $\flowvar$ now comes from 
the part of the torsion structure equation \eqref{torseq} 
projected into the normal space of $\map$, 
$\covder{t}\tangvec
= (\covder{x}\flowvec)_\perp
= \covder{x}\flowvec_\perp +\gtens(\flowvec_\perp,\covder{x}\tangvec)\tangvec
+ \flowvec_\parallel \covder{x}\tangvec$,
while the curvature structure equation \eqref{curveq}
and the tangential part of the torsion structure equation 
are used to express $\covder{t},\covder{x},\flowvec_\parallel$
in terms of $\cosympvar{}{},\flowvec_\perp$. 

There is a quite simple encoding of Hamiltonian operators 
in the frame components of these equations 
if the frame $\e{a}$ is chosen to be covariantly-constant
$\covder{x}\e{a}=0$,
\ie/ parallel transported, 
along the curve. 
(In a geometrical sense, 
whereas a parallel frame is completely adapted to the curve, 
a covariantly-constant frame is minimally adapted 
yet still determined entirely by geometrical considerations.)
To give some details:
the connection equation is simply
$\wtang{ab}{}=0$
while the curvature equation reduces to 
$\D{x}\wflow{ab}{} = -2\scalcurv \q{}{[a}\h{}{b]}$.
The torsion equation yields
$\D{x}\htang = \hperp{}{a}\q{a}{x}$
and so the flow equation becomes 
\EQ
\q{a}{t} 
= \cosympvar{a}{} -2\scalcurv \q{}{b}\Dinv{x}(\q{[a}{}\hperp{b]}{})
\endEQ
with 
\EQ
\cosympvar{a}{} 
= \D{x}\hperp{a}{} + \htang\q{a}{x} +\q{a}{}\hperp{}{b}\q{b}{x} .
\endEQ
Note the non-stretching condition on $\tangvec$ implies
$\cosympvar{a}{}\q{}{a}=\hperp{a}{}\q{}{a}=0$ as before,
but now $\q{a}{}$ is the Hamiltonian variable and obeys the constraint
$\q{a}{}\q{}{a}=1$. 
For ease of notation, 
$\q{a}{},\cosympvar{a}{},\hperp{a}{}$
will be written as $3$-component vectors
with $\cosympvec,\sympvec$ $\perp$ $\qvec$,
subject to $|\qvec|=1$.
A useful vector product identity 
\footnote[2]{Throughout, $\times$ denotes the standard cross-product on 
three-dimensional vectors.}
is 
$\vec{C}\hook(\vec{B}\wedge\vec{A}) 
= \vec{C}\times(\vec{B}\times\vec{A})$
in three dimensions. 
This leads to the operators
\EQ
\label{covframe-Iop}
\Iop = -\qvec\times\Dinv{x}(\qvec\times\ )
\endEQ
and 
\EQ
\label{covframe-Hop}
\Hop 
=\D{x} +\qvecder{x} \Dinv{x}(\qvecder{x}\cdot\ ) +\qvec(\qvecder{x}\cdot\ )
\endEQ
in terms of which the flow equation is given by 
\EQ
\qvecder{t}
= \scalcurv \Iop(\sympvec) + \Hop(\sympvec) . 
\endEQ
The following result can be proved 
by computations similar to \Ref{SandersWang2}
based on the framework presented in \Ref{Olver,Dorfman}. 
Details will be given elsewhere. 

{\bf Theorem~2}:\ {\it
$\Iop,\Hop$ are a pair of compatible Hamiltonian (cosymplectic) operators
and obey $\qvec\cdot\Iop=\qvec\cdot\Hop=0$,
with respect to the constrained Hamiltonian variable $\qvec(t,x)$. 
The inverse of $\Iop$ is a Hamiltonian symplectic operator
\EQ
\label{covframe-Jop}
\Jop=\Iop\inv = 
-\qvec\times\D{x}( 
(\qvec\times\ ) - \qvec\Dinv{x}((\qvec\times\qvecder{x})\cdot\ ) 
)
\endEQ
whose domain is defined on 
the $x$-jet space coordinates $\perp$ $\qvec$. 
\footnote[2]{Namely, 
the coordinate space 
$\{ (\qvecder{x},\qvecder{xx},\ldots)_\perp
=(\qvecder{x},\qvecder{xx}+|\qvecder{x}|^2 \qvec,\ldots) \}$
consisting of those vectors $\perp$ $\qvec$ 
derived from the differential consequences of 
$|\qvec|^2=1$.}
Then the flow equation on $\qvec$ takes the form
$\qvecder{t}
= \scalcurv \cosympvec' + \Rop(\cosympvec')$
in terms of the operator $\Rop=\Hop\circ\Jop$,
where $\cosympvec'=\Iop(\sympvec)$ 
is the inverse image of $\cosympvec=\Hop(\sympvec)$ 
under $\Rop$. 
}

As a result, 
$\Rop=\Hop\circ\Jop$ will be a hereditary recursion operator 
for commuting Hamiltonian vector fields 
$\cosympvec^{(k)}\cdot\Hfield{\qvec}$, 
$k=0,1,2,\ldots$,
starting from 
$\cosympvec^{(0)} = \qvecder{x}$, 
the infinitesimal generator of $x$-translations
in terms of the arclength $x$ along the curve $\map$. 
Likewise the adjoint operator $\Rop^*=\Jop\circ\Hop$ 
will be a hereditary recursion operator 
for involutive variational covector fields 
$\sympvec^{(k)}\cdot\coHfield{\qvec}$, 
$k=0,1,2,\ldots$,
related by $\sympvec^{(k)} = \Jop(\cosympvec^{(k)})$
and thereby starting from 
$\sympvec^{(0)}
=\Jop(\qvecder{x})
= \qvecder{xx}+|\qvecder{x}|^2 \qvec$. 
This hierarchy has a natural bi-Hamiltonian structure
$\cosympvec^{(k+1)} 
=\Iop(\sympvec^{(k+1)}) 
=\Hop(\sympvec^{(k)})$
with 
$\sympvec^{(k)} = -\delta \Ham{(k)}/\delta\qvec$
determining the Hamiltonians
$\Ham{}=\Ham{}(\qvec,\qvecder{x},\qvecder{xx},\ldots)$. 
Because of the constraint $|\qvec|=1$, 
the variational derivatives involve a Lagrangian multiplier term
$\lambda (|\qvec|-1)$ in the Hamiltonians
to enforce the constraint condition 
$\qvec\cdot\delta\Ham{}/\delta\qvec=0$. 
For instance, 
$\sympvec^{(0)}= \qvecder{xx}+|\qvecder{x}|^2 \qvec
=-\delta\Ham{(0)}/\delta\qvec$
determines 
$\Ham{(0)}= \frac{1}{2}|\qvecder{x}|^2 +\lambda (|\qvec|-1)$
where, after variational derivatives are evaluated, 
$\lambda=-|\qvecder{x}|^2$. 

The hierarchy 
$\cosympvec'=\cosympvec^{(k)}$, 
$k=0,1,2,\ldots$,
produces commuting bi-Hamiltonian flows on $\qvec$
given by constrained vector PDEs
\EQ
\qvecder{t} = \scalcurv\cosympvec^{(k)} + \cosympvec^{(k+1)} ,\quad
|\qvec|=1 . 
\endEQ
Their Hamiltonian structure looks like
\EQ
-\qvecder{t} 
= \Iop(\delta\Ham{(k,\scalcurv)}/\delta\qvec)
= \Hop(\delta\Ham{(k-1,\scalcurv)}/\delta\qvec)
\endEQ
where 
\EQ
\Ham{(k,\scalcurv)} = \scalcurv\Ham{(k)} + \Ham{(k+1)} 
\endEQ
are the Hamiltonians as before. 
These PDEs have an $\O{3}$-invariant form
on the $x$-jet space of $\qvec(t,x)$, 
where 
$\cosympvec' = \cosympvec^{(k)}(\qvec,\qvecder{x},\qvecder{xx},\ldots)$
and 
$\sympvec= \sympvec^{(k)}(\qvec,\qvecder{x},\qvecder{xx},\ldots)$ 
are covariant vector functions
given by explicit polynomial expressions.
Correspondingly, 
the curve $\map(t,x)$ undergoes a group-invariant non-stretching motion 
\EQ
\mapder{t}=
f(\mapder{x},\covder{x}\mapder{x},\covder{x}^2\mapder{x},\ldots) 
\endEQ
which is related to these vector expressions
by substitution of 
\EQ
\q{a}{}\e{a} = \mapder{x} ,\quad
\q{a}{x}\e{a} = \covder{x}(\q{a}{}\e{a}) = \covder{x}\mapder{x} ,\quad
\q{a}{xx}\e{a} = \covder{x}(\q{a}{x}\e{a}) = \covder{x}^2\mapder{x} , 
\endEQ
and so on, 
into the flow equation given by 
$f=\h{a}{}\e{a}$
with 
$\h{a}{} = \htang\q{a}{} + \hperp{a}{}$
and 
$\htang=\Dinv{x}(\hperp{}{a}\q{a}{x})$. 

The $0$ flow on $\qvec$ corresponds to 
the curve motion specified by 
the non-stretching mKdV map equation \eqref{mkdvmapeq}, 
which yields 
\EQ
\qvecder{t}
=\qvecder{xxx}+\frac{3}{2}(|\qvecder{x}|^2 \qvec)\downindex{x} 
+\scalcurv \qvecder{x} . 
\endEQ
This flow has the form of a vector mKdV potential equation
subject to the constraint $|\qvec|=1$. 
More generally, the $+k$ flow in the hierarchy 
is a higher-order mKdV constrained potential equation
arising from the mKdV map equation of order $3+2k$
in theorem~2
through the geometrical relation 
between the frame components $\sympvec$ of 
the curve motion $(\mapder{t})_\perp$
in the covariantly-constant frame and the parallel frame. 
This relation is expressed via the correspondence
\EQs
&&
\vredvec \leftrightarrow 
\covder{x}\mapder{x}
\leftrightarrow \qvecder{x} , 
\\
&&
\vredvecder{x} \leftrightarrow 
(\covder{x}^2\mapder{x})_\perp
\leftrightarrow \perpD{x}\qvecder{x} , 
\endEQs
and so on,
where (on the domain of $x$-jet space coordinates $\perp$ $\qvec$)
\EQ
\perpD{x}= \D{x} -\qvec \qvec\cdot\D{x} 
= \D{x} +\qvec (\qvecder{x}\cdot\ ) . 
\endEQ
Under this correspondence, 
the Hamiltonian operator \eqref{covframe-Hop}
and symplectic operator \eqref{covframe-Jop}
in the covariantly-constant frame
carry over respectively to the symplectic and cosymplectic operators 
\eqref{parframe-HJops}
in the parallel frame. 

{\bf Proposition~3}:\  
For any mKdV flow 
$\vredvecder{t}
= \sympredvec^{(\scalcurv)}(\vredvec,\vredvecder{x},\vredvecder{xx},\ldots)$
in the parallel frame
(where $\sympredvec^{(\scalcurv)}$ denotes a linear combination of
Hamiltonian vector fields 
$\sympredvec^{(k+1)}+\scalcurv\sympredvec^{(k)}$, $k=0,1,\ldots$),
the corresponding constrained potential mKdV flow
in the covariantly-constant frame
is explicitly given by 
$\qvecder{t}
= -\qvec\times\Dinv{x}( \qvec\times \sympvec^{(\scalcurv)}
(\qvecder{x},\perpD{x}\qvecder{x},(\perpD{x})^2\qvecder{x},\ldots) )$. 
This correspondence extends to the convective (traveling wave) flow
in the two frames. 

The converse correspondence is expressed via the augmented relations
\EQs
&&
\qvec 
\leftrightarrow \mapder{x} \leftrightarrow 
(\boldsymbol{0},1) ,
\\
&&
\qvecder{x} 
\leftrightarrow \covder{x}\mapder{x} \leftrightarrow 
(\vredvec,0) , 
\\
&&
\qvecder{xx} 
\leftrightarrow \covder{x}^2\mapder{x} \leftrightarrow 
\bigD{x} (\vredvec,0) , 
\endEQs
and so on,
using the derivative operator 
\EQ
\bigD{x} =\D{x} +(*\vredvec,0)\times . 
\endEQ
(A helpful identity relating $2$-component and $3$-component vectors here 
is $(\vredvec,0)\times(\boldsymbol{0},1) = (*\vredvec,0)$.)
This operator comes from the geometrical form of
the parallel frame $\e{a}$ in three dimensions as follows:
Write $(\e{a})_\perp \leftrightarrow \vece$
so $\e{a} =(\vece,\tangvec)$ is adapted to $\tangvec=\mapder{x}$. 
Then the conditions 
$\covder{x}(\e{a})_\perp=-\v{}{a}$
and $\covder{x}\tangvec=\flowvar$
for $\e{a}$ to be parallel \cite{Bishop}
are expressed by 
\EQ
\covder{x}(\vece,\tangvec) 
= (-\vredvec\otimes\tangvec,\flowvar)
= -(*\vredvec,0)\times (\vece,\tangvec)
\endEQ
where 
\EQ
\gtens(\flowvar,\vece)=\vredvec ,\quad
\gtens(\tangvec,\vece)=\gtens(\tangvec,\flowvar)=0 ,
\endEQ
with $\vredvec \leftrightarrow \v{}{a}$.
Hence a parallel frame has the invariant characterization 
$\bigcovder{x}\e{a}=0$ 
where $\bigcovder{x}$ is the covariant version of $\bigD{x}$. 

{\bf Proposition~4}:\  
Any constrained potential mKdV flow 
$\qvecder{t}
= \cosympvec^{(\scalcurv)}(\qvec,\qvecder{x},\qvecder{xx},\ldots)$
in the covariantly-constant frame
(where $\cosympvec^{(\scalcurv)}$ denotes a linear combination of
Hamiltonian vector fields 
$\cosympvec^{(k+1)}+\scalcurv\cosympvec^{(k)}$, $k=0,1,\ldots$)
has the corresponding form 
$\vredvecder{t}
= {\tilde\Hop}( \cosympredvec^{(\scalcurv)}
((\boldsymbol{0},1),(\vredvec,0),\bigD{x} (\vredvec,0),\ldots) )$
given by an mKdV flow,
using the Hamiltonian operator
${\tilde\Hop} = \D{x} + {*\vredvec} \Dinv{x}({*\vredvec}\cdot\ )$
from the parallel frame. 

The constraint $|\qvec|=1$ can be resolved by
the introduction of variables 
\EQ
\qvec=(\tangsg,\vecsg) ,
\endEQ
defined relative to any fixed unit vector $\unitvec$
by $\tangsg=\qvec\cdot\unitvec$ and $\vecsg=\qvec-\tangsg\unitvec$, 
satisfying 
\EQ
1=(\tangsg)^2+|\vecsg|^2 . 
\endEQ
Elimination of the scalar variable $\tangsg$ 
then leads to unconstrained potential mKdV flows
in terms of the $2$-component vector variable $\vecsg$. 
To within a convective term, 
the $0$ flow is given by 
\EQ
\vecsgder{t} 
= \vecsgder{xxx} +\frac{3}{2}( 
(|\vecsg|^2 (1-|\vecsg|^2)\inv (|\vecsg|\downindex{x})^2 
+ |\vecsgder{x}|^2 ) \vecsg )\downindex{x}
\endEQ
which is a vector generalization of the potential mKdV equation. 
In particular, 
for $\qvec=(\cos\sg,\sin\sg,0)$ lying in a plane, 
where $\tangsg=\cos\sg$, $\vecsg=(\sin\sg,0)$,
then $\sg$ satisfies the potential mKdV equation 
$\sgder{t}= \sgder{xxx} +\frac{1}{2}\sgder{x}^3$. 
This reduction has the geometrical meaning that 
the curve $\map(t,x)$ is restricted to lie on 
a totally-geodesic two dimensional subspace 
given by 
the sphere $\Sp{2}$ when $\scalcurv=1$ 
or the hyperboloid $\Hy{2}$ when $\scalcurv=-1$ 
or the plane $\Rnum{2}$ when $\scalcurv=0$
in the respective constant-curvature spaces 
$N=\Sp{3},\Hy{3},\Rnum{3}$. 

The same correspondence also applies to 
the curve motion specified by 
the non-stretching wave map equation \eqref{wavemapeq},
with $t$ conformally scaled so that $1=\gnorm{\mapder{t}}$,
which gives the $-1$ flow in the hierarchy. 
This flow is characterized by the kernel of the recursion operator 
$0=\cosympvec=\Rop(\cosympvec')=\Hop(\sympvec)$
in the covariantly-constant frame,
corresponding to the kernel of the symplectic operator in the parallel frame. 
Hence the $-1$ flow equation is given by 
$\qvecder{t} = \scalcurv \cosympvec' = \scalcurv \Iop(\sympvec)$
with $\sympvec$ determined by the geometrical equation 
$\cosympvec=\D{x}\hvec=0$
where the vector $\hvec=\htang\qvec + \sympvec$ is thus constant. 
This vector also satisfies
$|\hvec|=\gnorm{\mapder{t}}=1$,
which allows putting 
$\tangsg= \htang$, 
$\vecsg = \qvec-\htang\hvec$,
relative to $\hvec=\unitvec$. 
Then the tangential and normal parts 
of the $-1$ flow equation on $\qvec$ yield
nonlocal evolution equations
$\vecsgder{t} = -\scalcurv \tangsg\Dinv{x}\vecsg$
and 
$\tangsgder{t} = \scalcurv \vecsg\cdot\Dinv{x}\vecsg$, 
with the identity $1=(\tangsg)^2+ |\vecsg|^2$. 
Thus, after the scaling $t\rightarrow \scalcurv\inv t$ is made, 
the $2$-component vector variable $\vecsg$ satisfies 
the hyperbolic PDE
\EQ
\label{vecpotSGeq}
( \sqrt{(1-|\vecsg|^2)\inv}\ \vecsgder{t} )\downindex{x} = -\vecsg
%( \sqrt{1-|\vecsg|^2}^{-1}\ \vecsgder{t} )\downindex{x} = -\vecsg
\endEQ
which is a well-known vector generalization \cite{Bakas,PohlmeyerRehren}
of the SG equation. 
In particular, 
under the reduction 
$\tangsg=\cos\sg$, $\vecsg=(\sin\sg,0)$,
considered earlier 
where the curve $\map(t,x)$ lies on 
a totally-geodesic two dimensional subspace in $N$, 
the variable $\sg$ satisfies the ordinary SG equation. 

The bi-Hamiltonian operators in theorem~2 for the hierarchy of
potential mKdV flows are readily formulated by using 
the unconstrained variable $\vecsg$, 
providing a direct $2$-component vector generalization of the well-known 
bi-Hamiltonian structure of the SG hierarchy. 
Write 
$\symppot = \sympvec-\unitvec\cdot\sympvec\ \unitvec$, 
$\cosymppot = \cosympvec-\unitvec\cdot\cosympvec\ \unitvec$. 
Then we have 
\EQ
\vecsgder{t} = \scalcurv\Iop^\perp(\symppot) + \Hop^\perp(\symppot)
\endEQ
where the operators
\EQ
\Iop^\perp = 
\tangsg\Dinv{x}( \tangsg + \vecsg\tangsg\inv\vecsg\cdot\ )
-{*\vecsg}\Dinv{x}({*\vecsg\cdot}\ )
\endEQ
and 
\EQ
\Hop^\perp = 
\D{x} + \D{x}(\vecsg\Dinv{x}( 
(\vecsgder{x}\cdot\ ) - \tangsg\inv\tangsgder{x}\vecsg\cdot\ ))
\endEQ
are given by the normal parts of $\Iop,\Hop$ relative to $\unitvec$. 
The normal part of $\Jop$ relative to $\unitvec$ yields
an inverse operator for $\Iop^\perp$,
\EQs
\Jop^\perp = &&
\D{x} + \vecsg( (\vecsgder{x}\cdot\ ) 
- \tangsg\inv\tangsgder{x}\vecsg\cdot\, )
\\&&
-\tangsg{*\vecsg}\Dinv{x}( 
|\vecsg|(\tangsg |\vecsg|\inv)\downindex{x} ({*\vecsg}\cdot\ )
+ (\vecsgder{x}\cdot{*\vecsg}) \tangsg\inv|\vecsg|^{-2} \vecsg\cdot\ ) .
\nonumber
\endEQs
With respect to the unconstrained Hamiltonian variable $\vecsg(t,x)$,
it can be shown using computations similar to \Ref{SandersWang2} that
the operators $\Iop^\perp,\Hop^\perp$ 
are a compatible bi-Hamiltonian (cosymplectic) pair, 
while $\Jop^\perp$ is a Hamiltonian symplectic operator. 
Hence the operator $\Rop^\perp = \Hop^\perp \circ \Jop^\perp$
and its adjoint $\Rop^\perp{}^* = \Jop^\perp \circ \Hop^\perp$
are hereditary recursion operators
for the hierarchy of bi-Hamiltonian unconstrained vector potential mKdV flows
including the $-1$ flow given by the vector SG equation \eqref{vecpotSGeq}.

\subsection{ Left-invariant frames
and square-root operators }

Another choice of non-adapted moving frame $\e{a}$ is available
by utilizing the group-manifold structure that exists for
positive constant-curvature spaces in three dimensions. 
Recall that the compact semisimple Lie group $\SU{2}$ 
is isometric to the $3$-sphere. 
The isomorphism is expressed via left-invariant vector fields,
giving at any point on $N=\Sp{3}$ 
an identification of the tangent space
$T_{x}\Sp{3} \simeq \su{2} \simeq \so{3}$
with the Lie algebra of $\SU{2}$,
under which the Riemannian metric on $\Sp{3}$ is identified with
the Cartan-Killing metric of $\so{3}$. 
Any \on/ basis in $\so{3}$ thereby determines 
a left-invariant frame $\e{a}$ on the space $N=\Sp{3}$, 
satisfying the commutator property 
\EQ
[\e{a},\e{b}]
=2\sqrt{\scalcurv} \cross{c}{ab}\e{c}
\endEQ
and the curvature relation 
\EQ
\riem{a}{b}{\e{c},\e{d}} 
= \scalcurv \cross{e}{cd}\cross{b}{ea}
= 2\scalcurv \id{a[c}{}\id{d]}{b} ,
\endEQ
where $\cross{}{abc}=\cross{d}{ab}\id{cd}{}$
(Levi-Civita symbol)
represents the $\so{3}$ Lie-algebra structure constants,
and $\scalcurv>0$ is the curvature scalar of $N=\Sp{3}$. 
In contrast to a covariantly-constant frame,
a left-invariant frame is entirely adapted to the group-manifold geometry
of the space $N=\Sp{3}$
and so it is independent of any curves on this space. 
Nevertheless, such frames will be seen to naturally encode 
Hamiltonian operators in the frame components of 
the Cartan structure equations for 
group-invariant flows of curves $\map(t,x)$ on $N=\Sp{3}$.
These flows will be invariant under the group $\SO{3}$ of isometries that 
preserve the Lie algebra structure of $T_{x}\Sp{3}\simeq \so{3}$. 

Because of the non-adapted nature of the left-invariant frame, 
the components $\q{a}{}$ of the tangent vector $\tangvec$ along $\map$
are (non-constant) functions of $t,x$. 
So the most natural Hamiltonian variable will again be $\flowvar=\tangvec$
while $\flowvec_\perp = \flowvec-\flowvec_\parallel\tangvec$
and $\cosympvar{}{}=\covder{t}\tangvec$
continue to be 
the associated Hamiltonian vector and covector field variables.
The flow equation of motion has the same geometrical form as before, 
given by the part of the torsion structure equation \eqref{torseq}
projected into the normal space of $\map$. 
However, a crucial difference now is that 
the curvature structure equation \eqref{curveq}
becomes redundant since the connection $1$-form determined by
the left-invariant frame is just an algebraic expression
\EQ
\w{ab}{} = \cross{c}{ab}\e{c} . 
\endEQ
In detail:
the covariant derivative of the left-invariant frame 
in the tangent direction of the curve yields
\EQ
\covder{x}\e{a} 
= -\frac{1}{2}\sqrt{\scalcurv} \e{a}\btimes\tangvec
= -\sqrt{\scalcurv}\cross{c}{ab}\q{b}{}\e{c}
\endEQ
and in the flow direction yields
\EQ
\covder{t}\e{a} 
= -\frac{1}{2}\sqrt{\scalcurv} \e{a}\btimes\flowvec
= -\sqrt{\scalcurv}\cross{c}{ab}\h{b}{}\e{c} , 
\endEQ
giving the connection matrices
$\wtang{ab}{} = \sqrt{\scalcurv}\cross{}{abc}\q{c}{}$
and 
$\wflow{ab}{} = \sqrt{\scalcurv}\cross{}{abc}\h{c}{}$.
Here $\btimes$ denotes the $\so{3}$ Lie-algebra product on $T_{x}\Sp{3}$. 
Related to this structure, 
the frame components of 
$\auxcosympvar{}{}=\tangvec\btimes\flowvec$
will be seen to play the role of a Hamiltonian vector field.
The flow equation coming from the Cartan structure equations is
\EQ
\q{a}{t} = 
\D{x}\sympvar{a}{} +\htang \q{a}{x} +\q{a}{}\sympvar{}{b}\q{b}{x}
+2 \sqrt{\scalcurv} \auxcosympvar{a}{}
\endEQ
in terms of 
\EQ
\auxcosympvar{a}{}
= \cross{a}{bc} \q{b}{} \sympvar{c}{} ,
\endEQ
where 
$\D{x}\htang = \sympvar{}{a}\q{a}{x}$
is again obtained from the tangential part of the torsion equation,
and now 
\EQ
\cosympvar{a}{} 
= \D{x}\h{a}{} +\sqrt{\scalcurv} \auxcosympvar{a}{} . 
\endEQ
As before, 
$\q{a}{}\q{}{a} =1$
while 
$\auxcosympvar{a}{}\q{}{a} = \cosympvar{a}{}\q{}{a} 
= \sympvar{a}{}\q{}{a} =0$,
and the same $3$-component vector notation will be employed hereafter. 
Consequently,
the flow equation takes the Hamiltonian form 
\EQ
\qvecder{t} = \Hop(\sympvec) +2\sqrt{\scalcurv} \Iop(\sympvec)
\endEQ
where the operator
\EQ
\Hop 
=\D{x} +\qvecder{x} \Dinv{x}(\qvecder{x}\cdot\ ) +\qvec(\qvecder{x}\cdot\ )
\endEQ
is unchanged, 
but now 
\EQ
\Iop = \qvec\times
\endEQ
is an algebraic operator
(with the $\so{3}$ Lie-algebra product identified 
with the vector cross-product)
satisfying 
$\Iop^2 = \qvec\times(\qvec\times\ ) = -1 +\qvec (\qvec\cdot\ ) 
= -\idop_\perp$. 
Omitting details,
it can be proved that these operators enjoy the same properties
as the Hamiltonian operators \eqrefs{covframe-Hop}{covframe-Iop}
in the covariantly-constant frame. 

{\bf Theorem~5}:\ {\it
$\Iop,\Hop$ are a pair of compatible Hamiltonian (cosymplectic) operators
and obey $\qvec\cdot\Iop=\qvec\cdot\Hop=0$,
with respect to the constrained Hamiltonian variable $\qvec(t,x)$. 
The inverse of $\Iop$ is a Hamiltonian symplectic operator
$\Jop=\Iop\inv = -\qvec\times = -\Iop$
whose domain is defined on 
the $x$-jet space coordinates $\perp$ $\qvec$.
\footnote[2]{Namely, 
the coordinate space 
$\{ (\qvecder{x},\qvecder{xx},\ldots)_\perp
=(\qvecder{x},\qvecder{xx}+|\qvecder{x}|^2 \qvec,\ldots) \}$
consisting of those vectors $\perp$ $\qvec$ 
derived from the differential consequences of 
$|\qvec|^2=1$.}
Then the flow equation on $\qvec$ becomes
$\qvecder{t}
= 2\sqrt{\scalcurv} \cosympvec' + \Rop(\cosympvec')$
in terms of the operator $\Rop=\Hop\circ\Jop$,
where $\cosympvec'=\Iop(\sympvec)$ 
is the inverse image of 
$\cosympvec=\Hop(\sympvec)+\sqrt{\scalcurv}\Iop(\sympvec)$ 
under the shifted operator $\Rop'=\Rop +\sqrt{\scalcurv}$,
namely $\cosympvec=\Rop'(\cosympvec')$. 
}

On the $x$-jet space of $\qvec(t,x)$, 
it follows that $\cosympvec\cdot\Hfield{\qvec}$ has the geometrical meaning of
a Hamiltonian vector field
as does $\cosympvec'\cdot\Hfield{\qvec}$,
and $\sympvec\cdot\coHfield{\qvec}$ has the geometrical meaning of
a variational covector field. 
Hence 
$\Rop=\Hop\circ\Jop$ 
and its adjoint
$\Rop^*=\Jop\circ\Hop$ 
will be hereditary recursion operators
for respective hierarchies of 
commuting Hamiltonian vector fields $\cosympvec^{(k)}$
and involutive variational covector fields $\sympvec^{(k)}$,
$k=0,1,2,\ldots$,
related by $\sympvec^{(k)}=\Jop(\cosympvec^{(k)})$. 
Because $\Jop=-\Iop$ is purely an algebraic operator,
these recursion operators are like a ``square root'' of the ones
obtained from the covariantly-constant frame. 
The hierarchy starts at 
\EQ
\cosympvec^{(0)} = \qvecder{x} , 
\endEQ
describing infinitesimal $x$-translations
in terms of the arclength $x$ along the curve $\map$,
while 
\EQ
\sympvec^{(0)}
=\Jop(\qvecder{x})
= -\qvec\times \qvecder{x} . 
\endEQ
Next comes 
\EQ
\cosympvec^{(1)} 
= \Rop(\qvecder{x}) = -\Hop(\qvec\times \qvecder{x})
=  -\qvec\times \qvecder{xx} ,
\endEQ
and 
\EQ
\sympvec^{(1)}
= \Jop(-\qvec\times \qvecder{x})
= -(\qvecder{xx} +|\qvecder{x}|^2\qvec) ,
\endEQ
followed by
\EQ
\cosympvec^{(2)} 
=  -\qvecder{xxx}-\frac{3}{2} (|\qvecder{x}|^2\qvec)\downindex{x}
\endEQ
and 
\EQ
\sympvec^{(2)}
=  -\qvec\times\qvecder{xxx}
-\frac{3}{2} |\qvecder{x}|^2 \qvec\times\qvecder{x} .
\endEQ
In particular,
the vector fields $\cosympvec^{(0)},\cosympvec^{(2)}$, 
and so on for even $k$, 
along with the covector fields $\sympvec^{(1)},\sympvec^{(3)}$, 
and so on for odd $k$, 
exactly reproduce the hierarchy of 
potential mKdV Hamiltonian vector and covector fields
with respect to $\qvec(t,x)$ 
derived in the covariantly-constant frame. 

The full hierarchy in the left-invariant frame 
has a natural bi-Hamiltonian structure
$\cosympvec^{(k)} 
=\Iop(\sympvec^{(k)}) 
=\Hop(\sympvec^{(k-1)})$
with Hamiltonians
$\Ham{}=\Ham{(k)}(\qvec,\qvecder{x},\qvecder{xx},\ldots)$
determined by 
$\sympvec^{(k)} = \delta \Ham{(k)}/\delta\qvec$
subject to Lagrangian multipliers
enforcing the constraint condition 
$\qvec\cdot\delta\Ham{}/\delta\qvec=0$
like before. 
Thus, modulo the Lagrangian multipliers, 
\EQ
\Ham{(1)}=\frac{1}{2}|\qvecder{x}|^2
\endEQ
agrees with the bottom Hamiltonian in the hierarchy 
in the covariantly-constant frame,
while the next Hamiltonian in the hierarchy here is given by 
\EQ
\Ham{(2)}=-\frac{1}{2}\qvec\cdot(\qvecder{x}\times\qvecder{xx}) ,
\endEQ
and so on to all higher orders $k=1,2,\ldots$. 
But there is no Hamiltonian $\Hy{(0)}$
since any scalar triple product expressions that can be formed
just out of $\qvec$ and $\qvecder{x}$ are trivial. 

The entire hierarchy 
$k=0,1,2,\ldots$,
produces commuting flows 
\EQ
\qvecder{t} = 2\sqrt{\scalcurv}\cosympvec^{(k)} + \cosympvec^{(k+1)} , 
\endEQ
with $|\qvec|=1$. 
These flows describe constrained vector PDEs
having a bi-Hamiltonian structure
\EQ
\qvecder{t} 
= \Iop(\delta\Ham{(k,\scalcurv)}/\delta\qvec)
= \Hop(\delta\Ham{(k-1,\scalcurv)}/\delta\qvec)
\endEQ
given by the Hamiltonians 
\EQ
\Ham{(k,\scalcurv)} = 2\sqrt{\scalcurv}\Ham{(k)} + \Ham{(k+1)} . 
\endEQ
The form of the PDEs here is $\SO{3}$-invariant,
expressed by covariant vector functions
$\cosympvec'=\cosympvec^{(k)}(\qvec,\qvecder{x},\qvecder{xx},\ldots)$
and
$\sympvec=\sympvec^{(k)}(\qvec,\qvecder{x},\qvecder{xx},\ldots)$
consisting of explicit polynomials 
constructed using vector dot products and cross products
on the $x$-jet space of $\qvec(t,x)$. 
Compared to the larger $\O{3}$-invariance in the covariantly-constant frame,
invariance under reflections on $\qvec$ in the left-invariant frame 
is lost due to the explicit appearance of the cross-product operator. 

There is a corresponding hierarchy of group-invariant non-stretching motions
\EQ\label{mapfloweq}
\mapder{t}
= f(\mapder{x},\covder{x}\mapder{x},\covder{x}^2\mapder{x},\ldots)
\endEQ
on the underlying curve $\map(t,x)$.
These motions are obtained through substitution of 
\EQs
&&
\q{a}{}\e{a} = \mapder{x} , 
\\
&&
\q{a}{}\covder{x}\e{a}=0 ,
\\
&&
\q{a}{x}\e{a} = \covder{x}(\q{a}{}\e{a}) - \q{a}{}\covder{x}\e{a}
= \covder{x}\mapder{x} , 
\\
&&
\q{a}{x}\covder{x}\e{a} 
= \frac{1}{2}\sqrt{\scalcurv}\mapder{x}\btimes\covder{x}\mapder{x}
= -\frac{1}{2} \cross{c}{ab}\q{a}{x}\q{b}{} \e{c} ,
\endEQs
and so on, 
into the flow equation \eqref{mapfloweq} given by 
$f=\h{a}{}\e{a}$
with 
$\h{a}{} = \htang\q{a}{} + \hperp{a}{}$
and 
$\htang=\Dinv{x}(\hperp{}{a}\q{a}{x})$. 
The $0$ flow on $\qvec$
producing the vector PDE 
\EQ
\qvecder{t} = -\qvec\times\qvecder{xx} +2\sqrt{\scalcurv}\qvecder{x}
\endEQ
yields the curve motion 
\EQ
\label{groupmotionmapeq}
\mapder{t} = \mapder{x}\btimes\covder{x}\mapder{x} .
\endEQ
This motion defines a geometric map equation 
on the group-manifold $N=\Sp{3}$
and will be called the $\SO{3}$ {\it group-motion} equation for $\map$. 
A linear combination of this equation \eqref{groupmotionmapeq}
and the non-stretching mKdV map equation \eqref{mkdvmapeq}
gives the curve motion
determined by the $+1$ flow
\EQ
\qvecder{t} = 
-\qvecder{xxx} -\frac{3}{2}(|\qvecder{x}|^2 \qvec)\downindex{x} 
-2\sqrt{\scalcurv}\qvec\times\qvecder{xx} . 
\endEQ
Conversely the non-stretching mKdV map equation 
is given by a linear combination of the $0$ flow and $+1$ flow equations. 
A similar result holds for the hierarchy of higher order
non-stretching mKdV map equations when expressed
as vector PDEs for $\qvec$ in the left-invariant frame as follows. 

Through the geometrical meaning of $\sympvec$ and $\cosympvec$
as the respective frame components of 
the normal motion $(\mapder{t})_\perp$ of the curve 
and the evolution of its tangent vector $\covder{t}\mapder{x}$,
any constrained potential mKdV flow in the covariantly-constant frame
describes a non-stretching curve motion 
corresponding to a flow in the left-invariant frame 
(and conversely)
via replacing 
$\D{x} \leftrightarrow \covD{x}$
in the expressions for the Hamiltonian vector fields
$\cosympvec(\qvec,\qvecder{x},\qvecder{xx},\ldots)$
in the two frames,
where
\EQ
\covD{x} = \D{x} +\sqrt{\scalcurv}\qvec\times . 
\endEQ
This correspondence arises geometrically from 
the relation between the frame equations 
$0=\covder{x}\e{a}$ in the covariantly-constant case
and $0=(\covder{x}+ \sqrt{\scalcurv}\mapder{x}\btimes)\e{a}$ 
in the left-invariant case. 
The shifted recursion operator then has the interpretation as
a covariantized operator given by 
$\Rop \leftrightarrow \Rop'$
under replacing $\D{x} \leftrightarrow \covD{x}$
in the Hamiltonian operator $\Hop$. 

There is an analogous correspondence relating the flows 
in the left-invariant frame to the flows in the parallel frame
via the relations
\EQs
&&
\qvec 
\leftrightarrow \mapder{x} \leftrightarrow 
(\boldsymbol{0},1) ,
\\
&&
\covD{x}\qvec = \qvecder{x}
\leftrightarrow \covder{x}\mapder{x} \leftrightarrow 
(\vredvec,0) , 
\\
&&
\covD{x}\qvecder{x} 
\leftrightarrow \covder{x}^2\mapder{x} \leftrightarrow 
\bigD{x} (\vredvec,0) , 
\endEQs
and so on,
with 
\EQ
\bigD{x} =\D{x} +(*\vredvec,0)\times , 
\endEQ
while 
\EQ
\qvec\times \leftrightarrow * \leftrightarrow (\boldsymbol{0},1)\times ,
\endEQ
similarly to the correspondence 
for the case of the covariantly-constant frame. 

{\bf Theorem~6}:\ {\it 
For any flow 
$\qvecder{t}
= \cosympvec^{(\scalcurv)}(\qvec,\qvecder{x},\qvecder{xx},\ldots)$
in the left-invariant frame
(where $\cosympvec^{(\scalcurv)}$ denotes a linear combination of
Hamiltonian vector fields 
$\cosympvec^{(k+1)}+2\sqrt{\scalcurv}\cosympvec^{(k)}$, $k=0,1,\ldots$),
there is a geometrically corresponding flow in the parallel frame, 
$\vredvecder{t}
= {\tilde\Hop}^{(\scalcurv)}( \cosympredvec^{(\scalcurv)}
((\boldsymbol{0},1),(\vredvec,0),\bigD{x}^{(\scalcurv)}(\vredvec,0),\ldots) )
= \sympredvec^{(\scalcurv)}(\vredvec,\vredvecder{x},\ldots)$
in terms of the mKdV Hamiltonian operator
${\tilde\Hop} = \D{x} + {*\vredvec} \Dinv{x}({*\vredvec}\cdot\ )$,
where $\bigD{x}^{(\scalcurv)} = \bigD{x} - \sqrt{\scalcurv}*$
and ${\tilde\Hop}^{(\scalcurv)} = {\tilde\Hop} - \sqrt{\scalcurv}*$. 
Moreover, the shifted recursion operator in the left-invariant frame 
corresponds exactly to the square root of the mKdV recursion operator, 
\EQ
{\Rop'}^2 = -{\tilde\Hop}\circ{\tilde\Jop}
\endEQ
and 
\EQ
\Rop' = *{\tilde\Jop}= {\tilde\Hop}* , 
\endEQ
where ${\tilde\Jop} = \D{x} + \vredvec \Dinv{x}(\vredvec\cdot\ )$ 
is the mKdV symplectic operator. 
}

The Hodge-star operator $*$ here plays the role in the parallel frame
of an additional Hamiltonian operator ${\tilde\Iop}=-*$
which is compatible with ${\tilde\Hop}$
and corresponds to the symplectic operator $\Jop$ in the left-invariant frame.
Under this correspondence,
the $\SO{3}$ group-motion equation \eqref{groupmotionmapeq}
is given by the flow 
\EQ
\vredvecder{t} 
= *(\vredvecder{xx} +\frac{1}{2}|\vredvec|^2\vredvec +\scalcurv\vredvec)
\endEQ
with 
\EQ
\sympredvec=*\vredvec ,
\endEQ
which is outside the standard mKdV hierarchy. 
In particular, 
the square-root recursion operator $\Rop'$ generates 
an enlarged hierarchy of flows 
\EQ\label{enlarged}
\vredvecder{t} = {\Rop'}^k(*\vredvec)=\sympredvec^{(k)} ,\quad
k=0,1,\ldots, 
\endEQ
commuting with the mKdV flows. 
These flows \eqref{enlarged}
inherit a bi-Hamiltonian structure from their counterparts
in the left-invariant frame, 
\EQ
\vredvecder{t} 
= {\tilde\Hop}(\cosympredvec^{(k-1)}) = {\tilde\Iop}(\cosympredvec^{(k)})
\endEQ
in terms of the adjoint recursion operator
$\Rop^{'*} = {\tilde\Jop}*= *{\tilde\Hop}$, 
where 
\EQ
\cosympredvec^{(k)} = {\Rop'{}^*}^k(\vredvec) = *\sympredvec^{(k)} . 
\endEQ
Associated to this structure are 
commuting Hamiltonian vector fields
$\sympredvec^{(k)}\cdot\Hfield{\vredvec}$
and involutive covector fields
$\cosympredvec^{(k)}\cdot\coHfield{\vredvec}$, 
$k=0,1,\ldots$,
from which the Hamiltonians $\Ham{(k)}$ are determined by
\EQ
\delta\Ham{(k)}/\delta\vredvec = *\sympredvec^{(k)} =\cosympredvec^{(k)} . 
\endEQ
In detail, the hierarchy looks like:
\EQs
&&
\cosympredvec^{(0)} = \vredvec ,\quad
\Ham{(0)} = \frac{1}{2} |\vredvec|^2 ;
\\
&&
\cosympredvec^{(1)} = *\vredvecder{x} ,\quad
\Ham{(1)} = \frac{1}{2} \vredvec\cdot*\vredvecder{x} ;
\\
&&
\cosympredvec^{(2)} = 
-\vredvecder{xx}-\frac{1}{2} |\vredvec|^2 \vredvec ,\quad
\Ham{(2)} = \frac{1}{2} |\vredvecder{x}|^2 +\frac{1}{8}|\vredvec|^4 ;
\\
&&
\cosympredvec^{(3)} = 
-*(\vredvecder{xxx}+\frac{3}{2}|\vredvec|^2 \vredvec) ,\quad
\Ham{(3)} = 
\frac{1}{2} \vredvecder{x}\cdot*\vredvecder{xx}
-\frac{3}{8} |\vredvec|^2 \vredvec\cdot*\vredvecder{x} ,
\endEQs
and so on. 

This enlarged hierarchy contains the same $-1$ flow
as in the mKdV hierarchy, 
which arises directly from 
the kernel of the square-root recursion operator $\Rop'$.
In the left-invariant frame, this flow is given by 
$0=\cosympvec = \Hop(\sympvec)+\sqrt{\scalcurv}\Iop(\sympvec)
= \covD{x}\hvec$
where $\hvec = \sympvec + \htang\qvec$
is a covariantly-constant vector,
with $\D{x}\htang = \sympvec\cdot\qvecder{x}$. 
Note $\htang$ and $\sympvec$ are related through 
the conservation law $0=\D{x}|\hvec|^2$, 
with the consequence that $|\hvec|$ depends only on $t$
and hence satisfies $1=|\hvec|^2= |\sympvec|^2+\htang{}^2$
after $t$ is conformally scaled. 
The $-1$ flow equation produced on $\qvec$ is then given by 
\EQ
\qvecder{t} 
= \sqrt{\scalcurv}\cosympvec' = \sqrt{\scalcurv} \qvec\times\sympvec ,
\endEQ
or simply 
\EQ
\covD{t}\qvec=0
\endEQ 
so thus 
$\qvec$ is covariantly constant with respect to the covariant derivative
\EQ
\covD{t} = \D{t} +\sqrt{\scalcurv}\hvec\times. 
\endEQ
These equations in covariant-derivative form on $\qvec$ and $\hvec$
together are recognized to be 
a constrained vector $\su{2}$ chiral model
\cite{Mikhailov},
which is geometrically equivalent to 
the non-stretching wave map equation
$\cosympvar{}{} = \covder{t}\mapder{x}=0$ 
on the group-manifold $N=\SU{2} \simeq\Sp{3}$,
for the curve $\map(t,x)$ 
subject to $1=\gnorm{\mapder{x}}=\gnorm{\mapder{t}}$.

Interestingly, the wave map equation describing the $-1$ flow 
can be expressed as a vector hyperbolic equation on $\qvec(t,x)$,
via eliminating 
$\sympvec=-\sqrt{\scalcurv} \qvec\times\qvecder{t}$
and
$\htang = \sqrt{1-|\sympvec|^2} = \sqrt{1-\scalcurv\inv|\qvecder{t}|^2}$
to get
\EQ
\qvecder{tx} = 
-\qvecder{t}\cdot\qvecder{x} \qvec -\qvec\times(
\qvecder{t} +\sqrt{1-|\qvecder{t}|^2} \qvecder{x} )
\endEQ
after $\scalcurv$ is scaled out by 
$t\rightarrow t/\sqrt{\scalcurv}$, $x\rightarrow x/\sqrt{\scalcurv}$. 

Another formulation of the $-1$ flow equation is provided by 
a covariant-derivative vector SG equation 
derived as follows 
for the $2$-component vector variable $\sympredvec(t,x)$. 
First, differentiation of $0=\covD{t}\qvec$ by $\covD{x}$
followed by use of the commutator 
\EQ
[\covD{x},\covD{t}] 
= \scalcurv( \qvec\ {\hvec\cdot} - \hvec\ {\qvec\cdot} )
\endEQ
gives $0=\covD{t}\qvecder{x}-\scalcurv \sympvec$. 
Next, $\qvecder{x}$ can be eliminated in terms of $\htang$ and $\sympvec$
from $0= \covD{x}\hvec = \covperpD{x}\sympvec +\htang\qvecder{x}$
obtained by decomposition of $\hvec$ and $\covD{x}$
into tangential and normal parts relative to $\qvec$,
where the derivative operator 
\EQ
\covperpD{x} = \perpD{x} +\sqrt{\scalcurv}\qvec\times
\endEQ
satisfies $\qvec\cdot\covperpD{x}= \qvec\cdot\perpD{x}=0$. 
This operator has a natural meaning 
through the geometrical correspondence
\EQ
\vredvec \leftrightarrow \covder{x}\mapder{x} 
\leftrightarrow \qvecder{x} ,\quad
\vredvecder{x} \leftrightarrow (\covder{x}^2\mapder{x})_\perp 
\leftrightarrow \covperpD{x}\qvecder{x} ,  
\endEQ
and so on,
between the parallel frame and left-invariant frame. 
Finally, through the derivative operator
\EQ
\covperpD{t} = 
\perpD{t} +\sqrt{\scalcurv}(\sympvec\times\ + \htang\qvec\times)
\endEQ
which obeys $\qvec\cdot\covperpD{t}=0$, 
we have the equation
\EQ
-\covperpD{t}( \htang\inv \covperpD{x}\sympredvec ) = \scalcurv \sympredvec ,
\endEQ
where $\htang = \sqrt{1-|\sympredvec|^2}$. 

{\bf Theorem~7}:\ {\it 
In positive constant-curvature spaces 
$(N,\gtens)\simeq \Sp{3}\simeq \SU{2}$
in three dimensions, 
the group-invariant integrable flows of curves $\map(t,x)$
comprise an enlarged hierarchy 
whose even-order flows consist of the $\SO{3}$ group-motion equation
$$
\mapder{t} = \mapder{x}\btimes\covder{x}\mapder{x}
$$
and higher order analogs,
and whose odd-order flows consist of the non-stretching mKdV map equation 
$$
\mapder{t} 
= \covder{x}^2\mapder{x} 
+\frac{3}{2}\gnorm{\covder{x}\mapder{x}}^2 \mapder{x}
$$
and higher order analogs,
all subject to $\gnorm{\mapder{x}}=1$. 
The $-1$ flow is the non-stretching wave map equation
$$
\covder{t}\mapder{x}=0 ,\quad 
\gnorm{\mapder{x}}= \gnorm{\mapder{t}}=1 ,
$$
which is equivalent to the $\SU{2}$ chiral model. 
}

{\bf Corollary}:\ {\it
The $-1,0,+1$ flows can be reformulated as motions of
an arclength-parameterized embedded curve in $\Rnum{4}$
described by a $4$-component vector variable $\u(t,x)$.
This vector is constrained such that $|\u|=1$
so the curve motion is confined to the $3$-sphere $\Sp{3}\simeq\SU{2}$,
and in addition $|\uder{x}|=1$ 
so that the curve does not stretch (namely $x$ is arclength). 
Under the resulting identifications
\EQ
\map \leftrightarrow \u ,\quad
\covder{x} \leftrightarrow \D{x} -\u (\u\cdot\D{x}) , 
\endEQ
the wave map equation becomes the $\O{4}$ sigma model
\EQ
\uder{tx} =-\uder{t}\cdot\uder{x} \u ;
\endEQ
the mKdV map equation reduces to the model 
\EQ
\uder{t} = \uder{xxx}+\frac{3}{2} \uder{xx}\cdot\uder{xx} \uder{x}
\endEQ
up to a convective term;
and the $\SO{3}$ group-motion equation is given by the model
\EQ
\uder{t} = *(\u\wedge \uder{x}\wedge \uder{xx}) . 
\endEQ
Here $\cdot$ is the dot product, $\wedge$ the wedge product,
and $*$ the Hodge-star operator, on $\Rnum{4}$. 
}

\section{ Hamiltonian operators and integrable flows of curves 
in higher dimensions }

The derivation of bi-Hamiltonian operators 
from frame formulations of flows of non-stretching curves 
in three-dimensional spaces $\Sp{3}\simeq\SU{2},\Hy{3},\Rnum{3}$
generalizes to all higher dimensions
both for the case of constant-curvature spaces
and Lie-group spaces. 
As in three-dimensions, 
the flow of a curve $\map(t,x)$ in an $n$-dimensional Riemannian manifold
$(N,\gtens)$ for any $n\geq 2$
has associated to it the Cartan structure equations
expressing curvature
$[\covder{x},\covder{t}]
= \riem{}{}{\mapder{x},\mapder{t}}$
and torsion 
$\covder{x}\mapder{t} - \covder{t}\mapder{x} 
=[\mapder{x},\mapder{t}] =0$
in terms of \on/ frame vectors $\e{a}$
and connection $1$-forms $\w{ab}{}$
in the tangent space $T_{\map}N$. 
The same natural encoding of a compatible pair of Hamiltonian operators
contained in these equations is seen in all dimensions,
through the use of a geometrical choice of $\e{a}$ and $\w{ab}{}$
given by a covariantly-constant frame or a parallel frame
if $N$ is a constant-curvature space,
or a left-invariant frame 
if $N$ is a Lie-group space. 
For such frames the curvature matrix will be constant on $N$,
leading to a simple bi-Hamiltonian structure
based on the geometric variables given by 
the tangent vector $\tangvec = \mapder{x} = \q{a}{} \e{a}$
along $\map$,
the flow vector $\flowvec_\perp = (\mapder{t})_\perp = \hperp{a}{} \e{a}$
normal to $\map$,
and the principal normals 
$\flowvar = \covder{x}\tangvec =\v{a}{}\e{a}$
and $\cosympvar{}{} = \covder{t}\tangvec =\cosympvar{a}{}\e{a}$
associated with the tangential and flow directions of $\map$,
where $\gnorm{\tangvec}=1$ 
and $\gtens(\flowvar,\tangvec)=\gtens(\cosympvar{}{},\tangvec)=0$
for a non-stretching curve $\map$. 
Note the frame components of the principal normals 
$\flowvar,\cosympvar{}{}$ 
will be differential covariants of $\map$. 

\subsection{ Constant-curvature spaces }

Let $(N,\gtens)$ be 
an $n$-dimensional constant-curvature Riemannian manifold $(n\ge2)$,
with curvature scalar $\scalcurv$. 
The curvature matrix for any \on/ frame $\e{a}$ is simply
\EQ
\riem{a}{b}{\e{c},\e{d}} = 2\scalcurv \id{a[c}{}\id{d]}{b} .
\endEQ
Recall a parallel frame is adapted to $\map$ via
\EQ
\e{a} = ((\e{a})_\perp,\tangvec)
\endEQ
such that the covariant derivative of the normal vectors in the frame 
\EQ
\covder{x}(\e{a})_\perp = -\v{}{a}\tangvec
\endEQ
is tangential to $\map$
and the covariant derivative of the tangent vector 
\EQ
\covder{x}\tangvec = \v{a}{} (\e{a})_\perp
\endEQ
is normal to $\map$
(where, note, $\q{a}{}=\const$). 
The resulting Cartan structure equations 
carry over directly from three dimensions to $n$ dimensions
where $\v{a}{},\cosympvar{a}{},\sympvar{a}{}$ 
now are $n-1$-component vectors. 
Thus, the flow equation on $\v{a}{}$ retains the form
(using index-free notation)
\EQ
\vredvecder{t} = \Hop(\cosympredvec) + \scalcurv \Iop(\cosympredvec)
= \auxsympredvec +\scalcurv\sympredvec
\endEQ
where 
\EQ
\Hop = \D{x} + \vredvec\hook \Dinv{x}(\vredvec\wedge\ ) ,\quad
\Iop\inv=\Jop = \D{x} + \vredvec \Dinv{x}(\vredvec\cdot\ ) 
\endEQ
are, respectively, Hamiltonian cosymplectic and symplectic operators
and $\Rop = \Hop\circ\Jop$ is a hereditary recursion operator
as proved in \Ref{SandersWang2}. 
Here $\sympredvec=\Iop(\cosympredvec)$
and its image $\auxsympredvec=\Hop(\cosympredvec)$ under $\Rop$
have the roles of Hamiltonian vector fields
with respect to $\vredvec(t,x)$,
while the role of $\cosympredvec$ is a covector field. 
(Compared with \Ref{SandersWang2} 
the present formulation of these operators 
is manifestly $\O{n-1}$-invariant, 
employing just the vector dot product, interior and exterior vector products.
In particular, 
$\bi{C}\hook(\bi{B}\wedge\bi{A}) 
= (\bi{C}\cdot\bi{B})\bi{A} -(\bi{C}\cdot\bi{A})\bi{B}$.)

The resulting hierarchy of flows generated by $\Rop$ 
has the same structure as in three dimensions:
At the bottom
the $0$ flow is given by 
$\sympredvec = \vredvecder{x}$, 
%$\cosympredvec = \vredvecder{xx}+\frac{1}{2}|\vredvec|^2\vredvec =\Rop(\auxcosympredvec)$,
%$\auxcosympredvec = \vredvec$,
with $\vredvec$ seen to satisfy the vector mKdV equation
\EQ
\label{nvecmkdveq}
\vredvecder{t}= 
\vredvecder{xxx} +\frac{3}{2}|\vredvec|^2\vredvecder{x} 
+\scalcurv \vredvecder{x}
\endEQ
up to a convective term. 
The non-stretching curve motion of $\map$ produced by this flow 
is a corresponding mKdV map equation
\EQ
\label{mkdvmap}
\mapder{t} 
= \covder{x}^2\mapder{x} 
+\frac{3}{2}\gnorm{\covder{x}\mapder{x}}^2 \mapder{x}
\endEQ
with $\gnorm{\mapder{x}}=1$.
There is a $-1$ flow mapped into $\auxsympredvec=0$ 
under $\Rop$ in the hierarchy,
which is given by the kernel of $\Hop$
determined by $\cosympredvec=0$. 
This yields a nonlocal flow 
$\sympredvec = -\Dinv{x}(\htang\vredvec)$,
with the conservation law $\D{x}(|\sympredvec|^2+\htang{}^2)=0$
implying that the flow is conformally equivalent to one with uniform speed.
Thus, after a conformal scaling of $t$ is made, 
$\vredvec$ in this flow then satisfies the vector hyperbolic equation
\EQ
\label{nvechyperboliceq}
\vredvecder{tx} = -\sqrt{1-|\vredvecder{t}|^2} \vredvec . 
\endEQ
Equivalently, $\sympredvec$ obeys the vector SG equation
\EQ
( \sqrt{(1-|\sympredvec|^2)\inv}\ \sympredvec\downindex{x} )\downindex{t} 
= -\sympredvec . 
\endEQ
The corresponding curve motion of $\map$ is 
the non-stretching wave map equation 
\EQ
\label{wavemap}
\covder{t}\mapder{x}=0 ,\quad
\gnorm{\mapder{x}}=\gnorm{\mapder{t}}=1 ,
\endEQ
as seen from the geometrical meaning of $\cosympredvec$. 

A covariantly-constant frame provides a different formulation of
these main results, arising in terms of vector potential variables. 
Recall, since $\covder{x}\e{a}=0$,
the principal normal for the tangential direction of $\map$ 
for a covariantly-constant frame is given by 
$\v{a}{}=\q{a}{x}$ where $\q{a}{}\neq \const$. 
In the Cartan structure equations in $n$ dimensions,
$\q{a}{},\sympvar{a}{},\cosympvar{a}{}$ 
now are $n$-component vectors,
with $\q{a}{}$ having unit norm 
and with $\sympvar{a}{},\cosympvar{a}{}$ $\perp$ to $\q{a}{}$. 
Compared to three dimensions, 
vector cross products are replaced by interior and exterior products,
so thus the flow equation on $\q{a}{}$ becomes 
\EQ
\qvecder{t}
= \scalcurv \Iop(\sympvec) + \Hop(\sympvec)
= \scalcurv \auxcosympvec + \cosympvec
\endEQ
where 
\EQ
\Hop =
\D{x} +\qvecder{x} \Dinv{x}(\qvecder{x}\cdot\ ) +\qvec(\qvecder{x}\cdot\ )
\endEQ
as before,
while now 
\EQ
\Iop = -\qvec\hook\Dinv{x}(\qvec\wedge\ ) . 
\endEQ
The main results on these operators carry over to $n$ dimensions:
$\Iop,\Hop$ are a pair of compatible Hamiltonian (cosymplectic) operators
obeying $\qvec\cdot\Iop=\qvec\cdot\Hop=0$
with respect to the constrained variable $\qvec(t,x)$,
and so $\Jop=\Iop\inv$ defines a symplectic operator. 
(Note the domain of these operators consists of 
the $x$-jet space coordinates $\perp$ $\qvec$.
\footnote[2]{Namely, 
the coordinate space 
$\{ (\qvecder{x},\qvecder{xx},\ldots)_\perp
=(\qvecder{x},\qvecder{xx}+|\qvecder{x}|^2 \qvec,\ldots) \}$
consisting of those vectors $\perp$ $\qvec$ 
derived from the differential consequences of 
$|\qvec|^2=1$.})
The role of Hamiltonian vector fields is hence played by 
$\cosympvec=\Hop(\sympvec)$ 
and its inverse image $\auxcosympvec=\Iop(\sympvec)$ 
under the hereditary recursion operator $\Rop=\Hop\circ\Jop$. 
In dimensions $n>3$, however,
there is a noteworthy difference that an explicit simple expression 
for the inverse of $\Iop$ can no longer be derived. 

The hierarchy of flows in the parallel frame
corresponds to a hierarchy of constrained-potential flows 
in the covariantly-constant frame
as a result of the geometrical meaning of
the frame components $\sympvec$ for the normal motion 
$(\mapder{t})_\perp$ of the curve
as well as $\cosympvec$ for the evolution of the tangent vector 
$\covder{t}\tangvec$ of the curve. 
It is natural to express this correspondence
by first resolving the constraint $|\qvec|=1$
through splitting 
\EQ
\qvec=(\tangsg,\vecsg)
\endEQ
into a scalar variable $\tangsg$
and a $n-1$-component vector variable $\vecsg$
defined relative to any fixed unit vector $\unitvec$,
so 
\EQ
\tangsg=\sqrt{1-|\vecsg|^2} . 
\endEQ
This splitting yields the relations
\EQs
&&
\qvecder{x} = 
(-\tangsg\inv\vecsg\cdot\vecsgder{x},\vecsgder{x}) , 
\\
&&
\perpD{x}\qvecder{x} = 
(-\tangsg\inv\vecsg\cdot\perpD{x}\vecsgder{x},\perpD{x}\vecsgder{x}) , 
\endEQs
and so on,
where 
\EQ
\perpD{x} = \D{x} +\vecsg \tangsg((\tangsg\inv\vecsg)\downindex{x}\cdot\ )
\endEQ
which comes from a similar splitting of
$\perpD{x} = \D{x} +\qvec(\qvecder{x}\cdot\ )$. 
Hence the correspondence looks like
\EQ
\vredvec 
\leftrightarrow \covder{x}\mapder{x}
\leftrightarrow \vecsgder{x} ,\quad
\vredvecder{x} 
\leftrightarrow (\covder{x}{}^2\mapder{x})_\perp
\leftrightarrow \perpD{x}\vecsgder{x} , 
\endEQ
and so on. 
In particular, as in three dimensions,
the $0$ flow on $\vecsg$
is given by the vector potential mKdV equation
\EQ
\label{vecpotmkdveq}
\vecsgder{t} 
= \vecsgder{xxx} +\frac{3}{2}( 
(\tangsgder{x}{}^2 + |\vecsgder{x}|^2 ) \vecsg )\downindex{x}
\endEQ
up to a convective term,
while the $-1$ flow on $\vecsg$ reduces to the vector SG equation
\EQ
\label{vecSGeq}
( \tangsg\inv\ \vecsgder{t} )\downindex{x} = -\scalcurv\vecsg
\endEQ
after a conformal scaling on $t$ is used to put $\unitvec=\sympvec$. 
All these flows inherit a natural bi-Hamiltonian formulation 
which is obtained by splitting the operators $\Iop,\Hop$
relative to $\unitvec$.

{\bf Theorem~8}:\ {\it 
The flow equation on the unconstrained variable $\vecsg(t,x)$
takes the form 
\EQ
\vecsgder{t} 
= \scalcurv\Iop^\perp(\symppot) + \Hop^\perp(\symppot)
= \scalcurv\auxcosymppot + \cosymppot
\endEQ
given by 
\EQ
\Iop^\perp = 
\tangsg\Dinv{x}( \tangsg + \vecsg(\tangsg\inv\vecsg\cdot\ ) )
-\vecsg\hook\Dinv{x}(\vecsg\wedge\ )
\endEQ
and 
\EQ
\Hop^\perp = 
\D{x} + \D{x}(\vecsg\Dinv{x}( 
(\vecsgder{x}\cdot\ ) - \tangsg\inv\tangsgder{x}(\vecsg\cdot\ ) )) .
\endEQ
These operators $\Iop^\perp,\Hop^\perp$ 
are a compatible Hamiltonian pair
with respect to $\vecsg$,
where 
$\symppot = \sympvec-\unitvec\cdot\sympvec\ \unitvec$, 
$\cosymppot = \cosympvec-\unitvec\cdot\cosympvec\ \unitvec$
(and likewise for $\auxcosymppot$)
are Hamiltonian covector and vector fields, respectively. 
}

{\bf In summary}:
In any $n$-dimensional constant-curvature space $(N,\gtens)$
there is a hierarchy of bi-Hamiltonian flows of
non-stretching curves $\map(t,x)$,
where the $0$ flow is described by the mKdV map equation \eqref{mkdvmap}
and the $+k$ flow is a higher-order analog,
while the wave map equation \eqref{wavemap} describes a $-1$ flow 
that is annihilated by the recursion operator of the hierarchy.
In a parallel frame
the principal normal components along $\map$ for these flows
respectively satisfy
a vector mKdV equation \eqref{nvecmkdveq}
and a vector hyperbolic equation \eqref{nvechyperboliceq}. 
A covariantly-constant frame gives rise to 
potentials for the principal normal components along $\map$,
satisfying a vector potential mKdV equation \eqref{vecpotmkdveq}
and a vector SG equation \eqref{vecSGeq}, respectively. 

\subsection{ Hamiltonian operators in Lie group spaces }

Importantly, 
the idea of utilizing a non-adapted moving frame 
for non-stretching curve flows will now be carried over 
to Riemannian spaces $(N,\gtens)$
without the property of constant curvature,
specifically compact semisimple Lie group manifolds $N\simeq\G$. 
These spaces have a natural Riemannian structure \cite{KobayashiNomizu}
which is invariant under left-multiplication,
with the Riemannian metric $\gtens$ given by 
the bilinear Cartan-Killing form
in the Lie algebra $\liealg$ of the Lie group $\G$.
(Recall the Cartan-Killing form is 
nondegenerate when and only when $\G$ is semisimple, 
and positive when and only when $\G$ is compact.)
To proceed, let $(N\simeq\G,\gtens)$ be a $n$-dimensional 
(compact semisimple) Lie group manifold ($n>2$),
generalizing the three-dimensional case $N=\SU{2} \simeq \Sp{3}$
(cf. section~3.3). 
Any \on/ basis for the Lie algebra $\liealg\simeq T_{x}N$
provides a left-invariant \on/ frame $\e{a}$ on $N$
whose local structure group is $\G$ itself,
satisfying the commutator property 
\EQ
[\e{a},\e{b}]=\c{ab}{c}\e{c}
\endEQ
where $\c{ab}{c}=\c{[ab]}{c}$ denotes the Lie algebra structure constants. 
Note, 
the Jacobi relation is 
$\c{[ab}{e}\c{c]de}{}=0$
while semisimplicity is expressed by 
$\c{abc}{} = \c{[abc]}{}$,
where indices are raised and lowered by the Cartan-Killing form 
$\gtens(\e{a},\e{b}) = -\frac{1}{2}\c{ac}{d}\c{bd}{c} = \id{ab}{}$. 
The frame curvature matrix for $N$ is given by the algebraic expression 
\EQ
\riem{a}{b}{\e{c},\e{d}} = \frac{1}{4} \c{cd}{e}\c{ae}{b} . 
\endEQ
A distinguishing feature of a left-invariant frame is that 
this matrix is constant on $N$. 

The Cartan structure equations 
associated to the flow of a curve $\map(t,x)$ 
generalize from three dimensions, \ie/ $N=\SU{2}$, 
to $n$ dimensions, \ie/ $N=\G$,
with the structure constants $\c{ab}{c}$ in place of $\cross{c}{ab}$
and with the Lie-algebra bracket in place of a vector cross product. 
In particular, let $\adq\downupindices{a}{b} = \c{ca}{b}\q{c}{}$
denote the bracket on the $n$-component vector $\q{a}{}$. 
Note $\q{a}{}$ has the geometrical meaning of a potential for
the components $\v{a}{}=\q{a}{x}$ of the principal normal along $\map$. 
Hereafter it will be convenient to scale out a constant factor $\scalcurv$
from the curvature 
$\riem{a}{b}{\e{c},\e{d}} \rightarrow \scalcurv \c{cd}{e}\c{ae}{b}$
by putting 
$\c{ab}{c} \rightarrow 2\sqrt{\scalcurv}\c{ab}{c}$
for ease of comparison with the three dimensional case. 

Thus the flow equation on $\qvec$ (in index-free notation) 
is given by 
\EQ
\qvecder{t}
= \sqrt{\scalcurv} \Iop(\sympvec) + \Hop'(\sympvec)
= \sqrt{\scalcurv} \auxcosympvec + \cosympvec
\endEQ
where 
\EQ
\Iop=\adq
\endEQ 
is an algebraic operator
and 
\EQ
\Hop' = 
\covD{x} +\qvecder{x} \Dinv{x}(\qvecder{x}\cdot\ ) +\qvec(\qvecder{x}\cdot\ )
= \Hop+\sqrt{\scalcurv}\Iop
\endEQ
is a covariant version of the previous operator 
$\Hop 
= \D{x} +\qvecder{x} \Dinv{x}(\qvecder{x}\cdot\ ) +\qvec(\qvecder{x}\cdot\ )$
using the derivative 
\EQ
\covD{x} = \D{x} + \sqrt{\scalcurv}\adq . 
\endEQ
As in three dimensions,
the operators $\Hop$ (or $\Hop'$) and $\Iop$ 
are a compatible Hamiltonian pair
with respect to the constrained variable $\qvec(t,x)$,
on the domain of $x$-jet space coordinates $\perp \qvec$. 
So $\cosympvec$ and $\auxcosympvec$ represent Hamiltonian vector fields,
while $\sympvec$ represents a covector field. 
However in higher than three dimensions 
the operator $\Iop$ is no longer invertible
because of the fact that 
for any vector $\qvec$ in $\liealg$
the kernel of $\adq$ has a (nonzero) dimension at least equal to 
the rank of the Lie algebra $\liealg$,
and all semisimple Lie algebras of dimension $n>3$ have rank greater than $1$
(see \eg/ \cite{SagleWadle}).
Thus in the $n>3$ dimensional case, 
$\Iop\inv$ does not exist since $\Iop(\cvec{})=0$ 
for at least one nonzero vector $\cvec{}$ 
linearly independent of $\qvec$ in $\liealg$. 
This difficulty can be overcome by restricting the operators $\Hop,\Iop$ 
to a suitable smaller domain, 
which will be seen to necessarily introduce some nonlocality. 

Let $\Cq$ denote the centralizer of $\qvec$ in $\liealg$, 
namely the set of all vectors $\cvec{}$ commuting with $\qvec$ in $\liealg$,
\EQ
\adq\cvec{} =0 . 
\endEQ
Note that, for a given vector $\qvec$, 
the centralizer $\Cq$ is algebraically determined 
through the Lie algebra structure constants. 
Write $\perpCq$ for the orthogonal complement of $\Cq$ in $\liealg$
with respect to the Cartan-Killing metric,
so $\liealg = \Cq \oplus \perpCq$,
and write $\Cperpq$ for the orthogonal complement 
with respect to $\qvec$ in $\Cq$. 
The sets $\Cq,\Cperpq$ are Lie subalgebras of $\liealg$,
and the set $\perpCq$ is an invariant vector subspace 
in $\liealg$ under $\adq$. 
Let $\Pperpop,\Pop,\perpPop$ denote the projection operators in $\liealg$
onto $\Cperpq,\Cq,\perpCq$. 
Now decompose $\sympvec$ into the orthogonal parts 
$\Csympvec= \Pperpop(\sympvec)$, $\perpCsympvec= \perpPop(\sympvec)$. 

{\bf Proposition~9}:\
For all vectors $\cvec{}$ in $\Cperpq$, 
$\cvec{}\cdot\Iop=0$ 
while
$\cvec{}\cdot\Hop= 
-\cvecder{}{x}\cdot\perpPop 
- \cvecder{}{x}\cdot\qvec \Dinv{x}(\qvecder{x}\cdot\ )
+(\cvec{}\cdot\D{x})\Pperpop$. 

Hence 
$\Iop$ restricts to an operator in $\perpCq$ acting on $\perpCsympvec$ by 
\EQ
\Iop(\sympvec) = \adq\perpCsympvec := \Iop_\perp(\perpCsympvec) . 
\endEQ
Consequently, $\Iop_\perp$ is now invertible on $\perpCsympvec$,
with 
\EQ
\Iop_\perp^{-1}= \adqinv
\endEQ
where $\adop{\ }_\perp$ denotes the restriction of $\adop{\ }$ to $\perpCq$. 
Moreover, $\Hop$ can be adjusted to yield an operator in $\perpCq$
if the condition 
\EQ
\label{Hcondition}
\cvec{}\cdot\Hop(\sympvec)=0
\endEQ
for all vectors $\cvec{}$ in $\Cperpq$
is used to determine $\Csympvec$ in terms of $\perpCsympvec$
as follows:
Introduce an \on/ basis $\cvec{\mu}$ for the vector space $\Cperpq$,
so 
\EQ
\cvec{\mu}\cdot\cvec{\nu} = \id{}{\mu\nu} , 
\endEQ
and fix the $x$ dependence of the basis by the orthogonality property 
\EQ
\cvec{\mu}\cdot\cvecder{\nu}{x}=0 . 
\endEQ
Then the basis coefficients of the condition \eqref{Hcondition}
on $\Hop$ 
yield the equation
\EQ
\D{x}\hC{\mu} = 
\htang \cvec{\mu}\cdot\qvecder{x} + \cvecder{\mu}{x}\cdot\perpCsympvec
\endEQ
with 
\EQ
\D{x}\htang = 
\qvecder{x}\cdot\perpCsympvec + \hC{\mu} \cvec{\mu}\cdot\qvecder{x} ,
\endEQ
where $\hC{\mu} = \Csympvec\cdot\cvec{\mu}$ 
denotes the basis coefficients of $\Csympvec$. 
Hence $\Hop$ becomes an operator in $\perpCq$ acting on $\perpCsympvec$ by
\EQ
\Hop(\sympvec) = 
\D{x}( \perpCsympvec + \htang\qvec + \hC{\mu}\cvec{\mu} )
= \perpD{x}\perpCsympvec +\htang\perpqvecder{x} 
+\hC{\mu}\cvecder{\mu}{x}
:= \Hop_\perp(\perpCsympvec)
\endEQ
where 
\EQ
\perpD{x}= \perpPop(\D{x})
= \D{x}+\qvec (\qvecder{x}\cdot\ )+\ccovec{\mu} \cvecder{\mu}{x}
\endEQ
and 
\EQ
\perpqvecder{x} = \perpPop(\qvecder{x})
= \qvecder{x} + \ccovec{\mu} \qvec\cdot\cvecder{\mu}{x} .
\endEQ

To obtain the closest generalization of the results in three dimensions,
$\qvec$ should be restricted to obey the condition 
\EQ
\Pop(\qvecder{x})=0 ,
\endEQ
so thus $\qvecder{x}=\perpqvecder{x}$ will lie in the range of
the operators $\Hop_\perp,\Iop_\perp$. 
In this situation, 
$\htang$ and $\hC{\mu}$ are able to be explicitly determined
in terms of $\perpCsympvec$ by 
\EQ
\D{x}\htang = \qvecder{x}\cdot\perpCsympvec ,\quad
\D{x}\hC{\mu} = \cvecder{\mu}{x}\cdot\perpCsympvec . 
\endEQ
In addition, we have 
\EQ
\cvecder{\mu}{x} = \adqinv [\cvec{\mu},\qvecder{x}]
\endEQ
where the bracket denotes the Lie-algebra product.

The following result is a generalization of theorem~5:\ 
{\it 
$\Iop_\perp=\adq_\perp$
and 
$\Hop_\perp = 
\perpD{x} +\qvecder{x} \Dinv{x}(\qvecder{x}\cdot\ ) 
+\cvecder{\mu}{x} \Dinv{x}(\cvecder{\mu}{x}\cdot\ )$
are a pair of compatible Hamiltonian (cosymplectic) operators
obeying $\Pperpop(\Iop_\perp)= \Pperpop(\Hop_\perp)= 0$
with respect to the constrained variable $\qvec(t,x)$,
and $\Jop_\perp = \Iop_\perp^{-1} = \adqinv$
is a symplectic operator
(the domains of these operators consists of 
the $x$-jet space coordinates associated to $\perpCq$). 
The flow equation on $\qvec$ is given by 
$\qvecder{t}
= \sqrt{\scalcurv} \perpCauxcosympvec + \perpCcosympvec$
where 
$\perpCcosympvec=\Hop'_\perp(\perpCsympvec)$, 
$\perpCauxcosympvec=\Iop_\perp(\perpCsympvec)$
are Hamiltonian vector fields 
produced from the covector field $\perpCsympvec$. 
Here 
\EQ
\Hop'_\perp = 
\covperpD{x} +\qvecder{x} \Dinv{x}(\qvecder{x}\cdot\ ) 
+\cvecder{\mu}{x} \Dinv{x}(\cvecder{\mu}{x}\cdot\ )
= \Hop_\perp+\sqrt{\scalcurv}\Iop_\perp
\endEQ
is a covariant version of $\Hop_\perp$
with 
\EQ
\covperpD{x} = \perpD{x} +\sqrt{\scalcurv}\adq_\perp . 
\endEQ
Moreover 
$\Rop'=\Hop'_\perp\circ\Jop_\perp$ is a hereditary recursion operator. 
}

{\bf Remark}:
$N=\SU{2}\simeq S^3$ is 
the only compact semisimple Lie group of dimension $1<n\leq 3$.
For this group, 
the Lie bracket $\adq$ has an empty kernel for any (nonzero) vector $\qvec$
in the Lie algebra $\liealg=\su{2} \simeq \so{3}$
and hence the variables $\cvec{\mu}$ disappear.
Consequently, on the domain of $x$-jet space coordinates $\perp$ $\qvec$,
the operators $\Hop,\Iop$ are a compatible Hamiltonian pair
such that the inverse operator $\Jop=\Iop\inv$ exists and is 
a Hamiltonian symplectic operator,
and $\Rop=\Hop\circ\Jop$ is a hereditary recursion operator,
agreeing with the operators in theorem~5.

As in three dimensions,
$\Rop'$ generates a hierarchy of bi-Hamiltonian flows. 
At the bottom the $0$ flow is simply 
\EQ 
\perpCauxcosympvec = \qvecder{x} ,
\endEQ
while next the $+1$ flow is given by 
\EQ
\perpCauxcosympvec = \Rop'(\qvecder{x})
= \covD{x}( \adqinv\qvecder{x} 
+ \Dinv{x}(\cvecder{\mu}{x}\cdot\adqinv\qvecder{x}) \ccovec{\mu} )
\endEQ
which involves essential nonlocal terms 
in ($x$ derivatives of) the variables $\cvec{\mu}$. 
Accordingly, modulo such nonlocal terms,
the $0$ flow equation describes a natural generalization of
the $\SO{3}$ group-motion equation on $\qvec$ for the Lie group $\G$,
\EQ
\label{groupmotioneq}
(\qvecder{t})_{\rm local} = 
\perpD{x}(\adqinv\qvecder{x})
+2\sqrt{\scalcurv}\qvecder{x} .
\endEQ
Under this flow the curve $\map$ undergoes
a non-stretching motion given by the geometric map equation
\EQ
\label{groupmotionmap}
(\mapder{t})_\perp = \adinvop{\mapder{x}} \covder{x}\mapder{x}
\endEQ
with $\gnorm{\mapder{x}}=1$,
where $\perp$ denotes the projection with respect to $\perpPop$
in the tangent space $T_{x}N \simeq\liealg$ of the curve $\map$,
corresponding to 
\EQ
\perpCsympvec = \adqinv\qvecder{x} . 
\endEQ
In a similar way the $+1$ flow equation describes
a linear combination of the group-motion equation \eqref{groupmotioneq}
and a group-invariant generalization of the mKdV equation on $\qvec$, 
given by 
\EQ
(\qvecder{t})_{\rm local} =
\perpD{x}(\adqinv( \perpD{x}(\adqinv\qvecder{x}) ))
-\frac{1}{2}( |\adqinv\qvecder{x}|^2 \qvec )\downindex{x}
\endEQ
up to nonlocal terms. 
The curve motion of $\map$ produced by this flow 
is an analogous group-invariant mKdV map equation
which contains essential nonlocal terms,
corresponding to 
\EQ
(\perpCsympvec)_{\rm local} = 
\adqinv( \covD{x}(\adqinv\qvecder{x}) ) . 
\endEQ
Higher order (even and odd) flows are similar in form to the $0,+1$ flows. 

There is a $-1$ flow that is mapped into $\cosympvec=0$
under $\Rop'$ in the hierarchy. 
In this flow $\qvec$ and $\perpCsympvec$
satisfy the equations 
\EQ
0=\covD{t}\qvec=\covD{x}\hvec , 
\endEQ
which are equivalent to a constrained vector chiral model,
where $\hvec = \perpCsympvec + \htang\qvec +\hC{\mu}\cvec{\mu}$
obeys the conservation law $0=\D{x}|\hvec|^2$
and so $|\hvec|^2=1$ after $t$ is conformally scaled. 
Another formulation of the $-1$ flow equation is obtained by 
substituting 
\EQ\perpCsympvec = \adqinv\qvecder{t}
\endEQ
from the first equation into the second equation,
which then yields 
a vector hyperbolic equation on $\qvec(t,x)$, 
\EQ
\label{vechyperboliceq}
(\covD{x}\qvecder{t})_{\rm local} = 
[\qvecder{x},\adqinv\qvecder{x}]
\endEQ
up to certain nonlocal terms
involving $x$ derivatives of $\cvec{\mu}$. 
The corresponding curve motion of $\map$ is simply given by 
the non-stretching wave map equation
\EQ
\label{wavemapeq2}
\covder{t}\mapder{x}=0 ,\quad
1=\gnorm{\mapder{x}}=\gnorm{\mapder{t}} .
\endEQ

{\bf In summary}:
In $n$-dimensional Lie-group spaces $(N\simeq\G,\gtens)$
there is a hierarchy of bi-Hamiltonian flows of
non-stretching curves $\map(t,x)$,
subject to the geometric condition that the principal normal along $\map$
is orthogonal to the Lie-algebra centralizer in the normal space to $\map$, 
\ie/ $\gtens(\covder{x}\mapder{x},\Cmap)=0$
for all $\Cmap$ annihilated by $\adop{\mapder{x}}$. 
The hierarchy starts at the $0$ flow 
whose normal motion is described by 
the geometric group-motion map equation \eqref{groupmotionmap}
associated to $\G$,
while the wave map equation \eqref{wavemapeq2} describes a $-1$ flow 
that is annihilated by the recursion operator of the hierarchy.
In a left-invariant frame
the principal normal components along $\map$ for these flows
respectively satisfy
a vector group-motion equation \eqref{groupmotioneq}
and a vector hyperbolic equation \eqref{vechyperboliceq}.

\section{ Conclusion }

The main goal of this paper has been the study of 
various kinds of moving frames 
--- parallel, covariantly-constant, left invariant ---
in deriving bi-Hamiltonian operators and vector soliton equations
from flows of non-stretching curves in Riemannian manifolds.
A main insight, following the ideas of 
\Ref{SandersWang1,SandersWang2,AncoWang},
is that the bi-Hamiltonian structure is geometrically encoded
in the Cartan structure equations for the torsion and curvature
associated to a curve flow
whenever the frame curve matrix is everywhere constant on the manifold. 
The curve motions corresponding to the soliton equations 
determined by such a bi-Hamiltonian structure are found to be
geometric map equations,
in particular wave maps and mKdV/group-invariant 
analogs of Schr\"odinger maps. 

A unified treatment of constant-curvature manifolds and Lie-group manifolds
based on Riemannian symmetric spaces will be given in a forthcoming paper.

\ack
The author is supported by an N.S.E.R.C. grant. 
Jing Ping Wang is thanked for very fruitful discussions
in the early stage of this research. 
The referees are thanked for comments which have improved the paper.

\appendix
\section*{Appendix}

The following summary of Hamiltonian structure
is adapted from \Ref{Dorfman,Olver}.

Consider the jet space $\{(x,u,u_x,u_{xx},\ldots)\}$
of a scalar or vector variable $u(x)$.
Let $h\cdot\partial/\partial u$ be a vector field with component(s)
$h=h(x,u,u_x,u_{xx},\ldots)$,
and $\varpi\cdot du$ a covector field with component(s)
$\varpi=\varpi(x,u,u_x,u_{xx},\ldots)$,
where a dot denotes summation over any components. 
Write $\delta_h= {\rm pr}( h\partial/\partial u )$
denoting the variation (linearization) induced by $h$, 
\ie/ the prolonged action of the vector field;
write $\delta/\delta u$ for the variational derivative with respect to $u$.
Let $\theta$ denote a vertical uni-vector \cite{Olver}
dual to $du$, \ie/ $\theta\hook du=1$. 

A skew-adjoint operator $\Hop$ mapping $\varpi$ into $h$ is 
{\it Hamiltonian (cosymplectic)} iff
$$
0=\int\left( \theta\wedge \delta_{h} \Hop(\theta) \right) dx \quad
\eqtext{ for $h=\Hop(\theta)$} .
$$
This condition is equivalent to the vanishing of the Schouten bracket of
$\Hop$. 
A skew-adjoint operator $\Jop$ mapping  $h$ into $\varpi$ is 
{\it symplectic} iff
$$
0=\int\left( h_3\cdot \delta_{h_1} \Jop(h_2) + \eqtext{cyclic} \right) dx 
$$
for arbitrary $h_i$.
The operator $\Rop=\Hop\circ\Jop$ is a hereditary recursion operator on $h$
iff
$$
0= \Rop( \delta_{h_1}\Rop(h_2) - \delta_{h_2}\Rop(h_1) )
$$
for arbitrary $h_i$.

A (Hamiltonian) functional is an expression 
$\funct{H} = \int \Ham{}(x,u,u_x,u_{xx},\ldots) dx$.
The {\it Poisson bracket} with respect to a Hamiltonian operator $\Hop$ is
defined by 
$$
\{\funct{H},\funct{E}\}_\Hop
= \int\left( 
\Hop(\delta\funct{E}/\delta u)\cdot \delta\funct{H}/\delta u
\right) dx
$$
for any functionals $\funct{H},\funct{E}$.
This bracket is skew and obeys the Jacobi identity.
$h\cdot\partial/\partial u$ is a {\it Hamiltonian vector field}
if there exists a (Hamiltonian) functional $\funct{H}$ such that
$$
\delta_h \funct{E} = \{\funct{H},\funct{E}\}_\Hop
$$
holding for all functionals $\funct{E}$.
$\varpi\cdot du$ is a {\it variational (Hamiltonian) covector field}
if there exists a (Hamiltonian) functional $\funct{H}$ such that
$$
\varpi = \delta\funct{H}/\delta u .
$$

There is a canonical pairing between Hamiltonian vector fields
$h\cdot\partial/\partial u$
and variational covector fields 
$\varpi\cdot du$
via
$$
\varpi = \omega(h,\cdot)_\Jop
$$
where
$$
\omega(h_1,h_2)_\Jop = \int h_1\cdot \Jop(h_2) dx
$$
is a {\it symplectic $2$-form}. 
In particular, the operators $\Jop$ and $\Hop$ directly give mappings
$h\cdot\partial/\partial u \mapsto \Jop(h)\cdot du$
and
$\varpi\cdot du  \mapsto \Hop(\varpi)\cdot\partial/\partial u$.


\Bibliography{99}
\renewcommand{\i}{\char'020}

\bibitem{Lakshmanan}
M. Lakshmanan,
J. Math. Phys. 20, 1667--1672 (1979).

\bibitem{daRios}
R.L. Ricca, 
%Rediscovery of the da Rios equation,
Nature 352, 561--562, (1991).

\bibitem{Hasimoto}
H. Hasimoto,
J. Fluid Mech. 51, 477--485 (1972).

\bibitem{Shatah}
N.-H. Chang, J. Shatah, K. Uhlenbeck,
Comm. Pure Applied Math. 53, 590--602 (2000).

\bibitem{Polhmeyer}
K. Polhmeyer,
Comm. Math. Phys. 46, 207--221 (1976).

\bibitem{Lamb}
G.L. Lamb Jr.,
J. Math. Phys. 18, 1654--1661 (1977).

\bibitem{DoliwaSantini}
A. Doliwa, P.M. Santini,
Phys. Lett. A 185, 373--384 (1994). 

\bibitem{SandersWang1}
G. Mar\'\i\ Beffa, J. Sanders, J.-P. Wang,
J. Nonlinear Sci. 12, 143--167 (2002).

\bibitem{SandersWang2}
J. Sanders, J.-P. Wang,
Moscow Mathematical Journal 3, 1369--1393 (2003). 

\bibitem{Bishop}
R. Bishop,
Amer. Math. Monthly 82, 246--251 (1975).

\bibitem{AncoWang}
S.C. Anco, J.-P. Wang, 
in preparation (2005). 

\bibitem{Wang}
J.-P. Wang, 
in {\it Symmetry and Perturbation Theory}, 
eds. S. Abenda, G. Gaeta, S. Walcher
(World Scientific 2003). 

\bibitem{Guggenheimer}
H.W. Guggenheimer,
{\it Differential Geometry}
(McGraw Hill 1963).

\bibitem{WangSanders}
J. Sanders, J.-P. Wang,
J. Difference Equ. Appl. 12, 983--995 (2006). 

\bibitem{ChouQu1}
K.-S. Chou, C. Qu,
Physica D 162, 9--33 (2002).

\bibitem{ChouQu2}
K.-S. Chou, C. Qu,
Chaos, Solitons and Fractals 14, 29--44 (2002).

\bibitem{ChouQu3}
K.-S. Chou, C. Qu,
J. Nonlinear Sci. 13, 487--517 (2003).

\bibitem{ChouQu4}
K.-S. Chou, C. Qu,
Chaos, Solitons and Fractals 19, 47--53 (2004).

\bibitem{Dorfman}
I. Dorfman, 
{\it Dirac Structures and Integrability of Nonlinear Evolution Equations}
(Wiley 1993).

\bibitem{Olver}
P.J. Olver, 
{\it Applications of Lie Groups to Differential Equations}
(Springer, New York 1986).

%\bibitem{Boothby}
%W.M. Boothby,
%{\it An introduction to Riemannian manifolds and differential geometry}
%(Academic Press 1986)

\bibitem{KobayashiNomizu}
S. Kobayashi, K. Nomizu, 
{\it Foundations of Differential Geometry} Volumes I and II, 
(Wiley 1969).

\bibitem{MariBeffa}
G. Mar\'\i\ Beffa, 
Contemp. Math. 285, 29--38 (2001). 
%in {\it The Geometrical Study of Differential Equations}

\bibitem{SokolovWolf}
V.V. Sokolov and T. Wolf, 
J. Phys. A: Math. and Gen. 34, 11139--11148 (2001).

\bibitem{AncoWolf}
S. Anco, T. Wolf, 
J. Nonlinear Math. Phys. 12, 13--31 (2005); 
ibid. J. Nonlinear Math. Phys. 12, 607--608 (2005). 

\bibitem{Bakas}
I. Bakas, Q.-H. Park, H.-J. Shin, 
%Lagrangian formulation of symmetric space sine-Gordon models, 
Phys. Lett. B 372, 45--52 (1996). 

\bibitem{PohlmeyerRehren}
K. Pohlmeyer, K.-H. Rehren,
%Reduction of the two-dimensional ${\rm O}(n)$ nonlinear $\sigma $-model,
J. Math. Phys. 20, 2628--2632 (1979).

\bibitem{Mikhailov}
V.E. Zakharov, A.V. Mikailov,
Sov. Phys. JETP 47, 1017--1027 (1978).

\bibitem{SagleWadle}
A. Sagle and R. Walde, 
{\it Introduction to Lie Groups and Lie Algebras},
(Academic Press 1973). 




\endbib
\end{document}